\documentclass[a4paper,11pt]{article}
\pdfoutput=1

\usepackage{amsmath}
\usepackage{graphicx}
\usepackage{amsfonts}
\usepackage{amssymb}

\newtheorem{theorem}{Theorem}[section]
\newtheorem{rema}{Remark}[section]

\newtheorem{defi}[rema]{Definition}
\newtheorem{lemma}[theorem]{Lemma}
\newtheorem{corol}[theorem]{Corollary}

\newcommand{\bc}{\begin{center}}
\newcommand{\ec}{\end{center}}
\def\ba#1{\begin{array}{#1}\displaystyle}
\newcommand{\ea}{\end{array}}

\newcommand{\beq}{\begin{equation}}
\newcommand{\eeq}{\end{equation}}
\newcommand{\beqa}{\begin{eqnarray}}
\newcommand{\eeqa}{\end{eqnarray}}
\newcommand{\no}{\nonumber}
\newcommand{\n}{\nonumber\\}
\newcommand{\bi}{\begin{itemize}}
\newcommand{\ei}{\end{itemize}}

\def\lt#1{\left#1}
\def\rt#1{\right#1}
\def\t#1{\tilde{#1}}

\def\frc#1#2{\frac{#1}{#2}}
\newcommand{\p}{\partial}

\newcommand{\Pexp}{{\cal P}\exp}

\newcommand{\bra}{\langle}
\newcommand{\ket}{\rangle}
\newcommand{\Z}{{\mathbb{Z}}}
\newcommand{\N}{{\mathbb{N}}}
\newcommand{\R}{{\mathbb{R}}}
\newcommand{\C}{{\mathbb{C}}}

\newcommand{\dd}{\mathrm{d}}
\newcommand{\ii}{{\mathrm{i}}}

\newcommand{\Or}{{\cal O}}

\newcommand{\ep}{\epsilon}

\newcommand{\Tr}{{\rm Tr}}

\newcommand{\halmos}{\rule{1ex}{1.4ex}}
\newcommand{\eproof}{\hspace*{\fill}\mbox{$\halmos$}}
\newcommand{\proof}{{\em Proof.\ }}

\DeclareMathOperator{\ad}{ad}

\newcommand{\1}{{\bf 1}}
\newcommand{\obs}{{\cal O}}

\DeclareMathOperator{\supp}{{\rm supp}}
\def\siz#1{{|#1|}}
\DeclareMathOperator{\dist}{{\rm dist}}
\DeclareMathOperator{\diam}{{\rm diam}}
\newcommand{\iso}{\iota}
\newcommand{\obsd}{{\cal N}}
\newcommand{\obsn}{{\cal L}}
\newcommand{\obsh}{{\cal H}}
\newcommand{\obsq}{{\cal Q}}
\newcommand{\obsc}{{\cal A}}
\newcommand{\dbra}{{\langle\langle}}
\newcommand{\dket}{{\rangle\rangle}}
\newcommand{\obsdd}{\widehat{\cal N}}
\newcommand{\obsdn}{\widehat{\cal L}}
\newcommand{\obsdh}{\widehat{\cal H}}
\newcommand{\obsdq}{\widehat{\cal Q}}
\newcommand{\dens}{{\frak D}}
\newcommand{\ddens}{\widehat{\frak D}}
\newcommand{\map}{{\frak M}}
\newcommand{\qmap}{{\frak q}}
\newcommand{\sta}{{\rm sta}}
\newcommand{\rep}{{\frak L}}
\newcommand{\di}{{\tt D}}
\newcommand{\are}{{\tt S}}
\newcommand{\ball}{{\tt B}}

\begin{document}

\begin{titlepage}

\begin{center}
{\Large {\bf Thermalization and pseudolocality\\ in extended quantum systems}

\vspace{1cm}

Benjamin Doyon}

Department of Mathematics, King's College London\\
Strand, London WC2R 2LS, U.K.\\

\end{center}

\vspace{1cm}

\noindent Recently, it was understood that modified concepts of locality played an important role in the study of extended quantum systems out of equilibrium, in particular in so-called generalized Gibbs ensembles. In this paper, we rigorously study pseudolocal charges and their involvement in time evolutions and in the thermalization process of arbitrary states with strong enough clustering properties. We show that the densities of pseudolocal charges form a Hilbert space, with inner product determined by thermodynamic susceptibilities. Using this, we define the family of pseudolocal states, which are determined by pseudolocal charges. This family includes thermal Gibbs states at high enough temperatures, as well as (a precise definition of) generalized Gibbs ensembles. We prove that the family of pseudolocal states is preserved by finite time evolution, and that, under certain conditions, the stationary state emerging at infinite time is a generalized Gibbs ensemble with respect to the evolution dynamics. If the evolution dynamics does not admit any conserved pseudolocal charges other than the evolution Hamiltonian, we show that any stationary pseudolocal state with respect to this dynamics is a thermal Gibbs state, and that Gibbs thermalization occurs. The framework is that of translation-invariant states on hypercubic quantum lattices of any dimensionality (including quantum chains) and finite-range Hamiltonians, and does not involve integrability.

\vfill

{\ }\hfill 
\today

\end{titlepage}

\section{Introduction}

The physics of quantum systems out of equilibrium has received a large amount of attention recently. In this context, a problem of great interest is that of thermalization: Can a quantum system reach a thermalized state under unitary evolution after a long time? What conditions are there on the initial state and on the evolution Hamiltonian for this to happen? The recent reviews \cite{PSS11,Y11,GE15,EFG15} offer clear discussions of these and related questions.

In order to avoid quantum recurrence, the system should be composed of infinitely-many degrees of freedom. Consider homogeneous, extended systems, where the finite-dimensional local degrees of freedom lie on a regular lattice and interact locally in a uniform fashion (so that the energy is extensive with the volume of finite sub-lattices). In this context, the $C^\star$ algebra $\obsc$ of observables is a completion of the space of local observables, and states are normalized positive linear functionals on $\obsc$ representing their averages. The dynamics $\tau_t(A)=e^{\ii Ht} A e^{-\ii Ht}$, where $H$ is the Hamiltonian, has a unique, unambiguous meaning for any $A\in\obsc$. Thermal Gibbs states $\omega_\beta^{\rm th}$ at temperature $\beta^{-1}$ with respect to $H$ may be defined in various fashions: the Gibbs condition, the maximum entropy principle, the $(\tau,\beta)$-KMS condition; these are equivalent if, for instance, $H$ only has local interactions as assumed here \cite[Cor. 6.2.37]{BR1}, and $\omega_\beta^{\rm th}$ is unique for large enough temperatures \cite[Cor. 6.2.45]{BR1} (and in one dimension \cite{Araki69}). See \cite{Ruelle,Ta,Israel,BR1,Simon,Sak93,JP,AJPP,JOPP}
for expositions on the subject (and \cite{PY94,PY95} for some extensions to infinite-dimensional local spaces). A natural question of thermalization may then be formulated as follows\footnote{We separate the question of {\em equilibration}, the modes of existence of the long-time limit, from that of {\em thermalization}, the reaching of a thermal state, see for instance \cite{GE15}.}: if an evolution dynamics $\tau_t$, initial state $\omega$ and family of observables ${\cal F}\subset\obsc$ are such that the long-time limit $\lim_{t\to\infty} \omega(\tau_t(A))$ exists for all $A\in{\cal F}$, is there a thermal Gibbs state $\omega_\beta^{\rm th}$, $\beta>0$ (a $(\tau,\beta)$-KMS state) such that $\lim_{t\to\infty} \omega(\tau_t(A)) = \omega_\beta^{\rm th}(A)\;\forall\;A\in{\cal F}$? There is a large body of results related to this question, for instance results based on the eigenstate thermalization hypothesis \cite{JenShan85,deu,sre,Tas98,rigol1}, and rigorous thermalization theorems under other conditions and for certain families of initial states \cite{RGE12,SKS13,MAMW15}.

The above might be too strong: Gibbs states obtained after large times may naturally involve any other local charges $Q$ (with densities that are local observables) that are conserved by the dynamics $H$, with their own chemical potentials, formally with (un-normalized) density matrices of the form $e^{-\beta(H-\mu Q - \ldots)}$. Clearly, then, the question of thermalization is strongly affected by the local conservation laws afforded by the dynamics.

It is straightforward to extend the idea to situations with infinitely-many conserved local charges $Q_i$ in involution, as was envisaged by Jaynes early on \cite{jaynes1,jaynes2}. Such infinite families are present in integrable models, where powerful exact methods are available. The resulting states, formally with density matrices $e^{-\sum_{i=1}^{\infty}\beta_i Q_i}$, are referred to as generalized Gibbs ensembles (GGEs) \cite{GGE1,rigol2}, and the question of ``generalized thermalization'' has been studied intensively in recent years. The idea is to fix the chemical potentials $\beta_i$ sequentially, in truncations of the infinite series $\sum_{i=1}^{\infty}\beta_i Q_i$ and in finite-volume regularizations, by using the averages of conserved charges in the initial state. This should provide a converging method for predicting the stationary state, and it has been verified that such states indeed appear to be reached after long times in many examples \cite{GGE1,rigol2,GGE2,GGE3,GGE4,GGE5,GGE51,GGE6,GGE7,GGE8,GGE9,GGE9-1,GGE10,GGE11,GGE12,GGE13,GGE14,SC14,Fa14,MS15,Ca15,GKFGE}. However, in some cases, this truncation method, although converging well, does {\em not} lead to good approximations of infinite-time stationary states \cite{WNB14,PMW14,MPTW15}. As proposed in \cite{EMP15,IlieGGE}, one must include conserved quasi-local charges, whose densities have exponentially decaying tails. It was numerically shown in \cite{IlieGGE}, using the Bethe ansatz and comparing with results \cite{WNB14,BWF14} obtained with the quench action formalism of Caux and Essler \cite{CE13}, that a truncation involving conserved quasi-local charges leads to a good approximation of the stationary state in certain cases where the use of local charges failed\footnote{More precisely, this conclusion holds under the assumption that the quench action formalism provides the correct large-time steady state, but this is a statement about which there is little doubt in the community.}. Quasi-local charges were originally proposed by Ilievski and Prosen \cite{PI13,IP12} in a different context, and were calculated in various integrable models \cite{Pro14,PPSA14,IMP15,ZMP16} and explicitly related to Bethe strings \cite{IQDB}. They are special examples of pseudolocal charges, defined by their volume-scaling properties and first introduced by Prosen \cite{Pr98,Pr99,Pr11}.

Formal density matrices $e^{-\sum_{i=1}^{\infty}\beta_i Q_i}$, where $Q_i$ are conserved charges, are {\em a priori} ill-defined and do not clearly characterize GGEs as linear functionals on $\obsc$. For instance, in what sense does the series $\sum_{i=1}^{\infty}\beta_i Q_i$ converge? What locality properties does the resulting conserved charge possess? Can it be a non-local conserved charge? In the latter case, the statement of generalized thermalization might be conceptually hollow. Numerical and analytical results thus far obtained do suggest meaningful convergence properties. Also, GGEs may be naturally defined \cite{GE15} in finite quantum systems using entropy maximization. Alternatively, it might be possible to define GGEs as equilibrium (say KMS) states with respect to Hamiltonians with nonlocal interaction potentials lying in an appropriate Banach space, see for instance \cite{Israel}. Yet, a full understanding of (generalized) thermalization and the role of extended concepts of locality is largely missing.

In this paper, we propose a new rigorous approach that sheds light on these problems. We consider general quantum states satisfying strong enough clustering properties that guarantee the existence and finiteness of thermodynamic susceptibilities. Within this context, we prove that pseudolocal charges (a refinement, and generalization to arbitrary dimensions, of the concepts developed in \cite{Pr98,Pr99,Pr11,PI13,IP12}) are in bijection with the countable-dimensional Hilbert space induced by Cauchy completing the positive semidefinite sesquilinear form built out of susceptibilities. This Hilbert space can be seen as the space of the pseudolocal charges' densities. The fact that pseudolocal charges form a Hilbert space was noted in \cite{Pr98,Pr99}, and our construction transfers this structure to their densities. A series of the type $\sum_i \beta_i Q_i$ may hence be interpreted as a basis decomposition of a single pseudolocal charge.

We then define a family of {\em pseudolocal states}. These are clustering states connected to the infinite-temperature state by paths whose tangents are determined by pseudolocal charges (conserved or not by any dynamics). They can be seen, formally, as states with density matrices written as path-ordered exponentials $\Pexp \int_0^s\,\dd u\,Q_u$ where $Q_u$ are pseudolocal charges. This family contains, and generalizes, high-temperature thermal Gibbs states ($Q_u=-H$ for all $u$) associated to local Hamiltonians $H$. We prove that the family of pseudolocal states is preserved under finite time evolutions with any finite-range (homogeneous) dynamics, showing that it is, in some sense, a natural family of states.

If the pseudolocal charges $Q_u$ are almost everywhere (along the path)  conserved by some fixed local dynamics, then the pseudolocal state is said to be a GGE with respect to this dynamics. This gives a precise definition of GGEs (note that this definition accounts for the possibility that there be conserved charges that generate noncommuting flows). We prove that, under certain conditions on the existence of the long-time limit, any pseudolocal state reaches a GGE with respect to the evolution dynamics. The conditions include the existence of the long-time limit of dynamical susceptibilities. This generalized thermalization result holds independently of integrability, the latter only affecting the family of conserved pseudolocal charges available and thus the manifold of final states. In particular, if no conserved pseudolocal charges are available other then the evolution Hamiltonian, we show that Gibbs thermalization occurs. Generalized thermalization is our main result, and justifies, we believe, the introduction of the notion of pseudolocal states.

All results are established within the framework of quasi-local $C^\star$ algebras on hypercubic quantum lattices of any finite dimension (including quantum chains), and we expect the results to hold on any regular lattice of finite dimension. We do not have, however, the full connection with established $C^\star$-algebra concepts, as we do not know yet the relation between GGEs and the equilibrium states associated to (generically nonlocal) interaction potentials.

The paper is organized as follows. In Section \ref{sectov} we provide an overview of the definitions and main results. In Section \ref{sectalg}, we specify the mathematical framework. In Section \ref{sectclust}, we define and study the clustering properties of states, that are essential for the rest of the paper. In Section \ref{sectquasi}, we define the concepts of pseudolocal sequences and pseudolocal charges, and prove theorems relating to these. In Section \ref{secttime}, we use pseudolocal charges in order to study time evolution and prove the  reaching of a GGE (Theorem \ref{theoGGE}) and re-thermalization (Corollary \ref{theore}). Finally, in Section \ref{sectconclu} we provide interpretations and conclusions.

\section{Overview of main results}\label{sectov}

Let $\di\geq 1$ be an integer and consider the regular $\di$-dimensional infinite hypercubic lattice $\Z^\di$, each site supporting a finite-dimensional space of square matrices. Consider translation-invariant states $\omega$ on this quantum lattice. See Section \ref{sectalg} for the precise framework.

\subsection{Hilbert space and pseudolocal charges}

Let $A$ and $B$ be local observables (observables supported on a finite number of sites, acting like the identity on other sites), the space of which we denote $\obs$, a dense subspace of the $C^\star$ algebra $\obsc$. Consider symmetrized susceptibilities, which are total volume integrals of symmetrized, connected correlation functions,
\beq\label{ovresp}
	\dbra A,B\dket_\omega
	= \sum_{x\in\Z^\di} \lt(\frc12\omega(\{A_x^\star, B\})
	-\omega(A^\star)\omega(B)\rt).
\eeq
Here $A_x^\star$ is the displacement of the conjugate $A^\star$ of $A$ by $x\in\Z^\di$, and $\{\cdot,\cdot\}$ is the anti-commutator. In order for the series \eqref{ovresp} to exist, we assume $\omega$ has strong enough clustering properties at large distances: $\omega(AB)\approx \omega(A)\omega(B)$ up to terms that decay as $\dist(A,B)^{-p}$ ($p>\di$) with the distance (see Definitions \ref{deficlust} and \ref{defisiz} and Theorem \ref{propoconv}). The function $\dbra\cdot,\cdot\dket_\omega$ is a positive semidefinite sesquilinear form (Theorem \ref{propopos}), and by standard arguments (see the Gelfand-Naimark-Segal construction \cite{Gelfand,Segal}), we may Cauchy complete the quotient of the space of local observables by the subspace on which $\dbra\cdot,\cdot\dket_\omega$ is degenerate, thus obtaining a Hilbert space $\obsdh_\omega$.

We define pseudolocality (Definition \ref{defiquasi}) following closely \cite{Pr98,Pr99,Pr11} and \cite{PI13,IP12}, but generalizing to clustering states and to any dimension, and with other adjustments. We will see that the above Hilbert space is connected to the space of pseudolocal charges. Consider sequences of local observables $Q_n$ supported on ``balls'' $\ball(n)\subset \Z^\di$ centered at the origin and of radii $n$. Assume $\omega(Q_n)=0$. A pseudolocal sequence   $Q=\{n\mapsto Q_n\}$ with respect to $\omega$ is defined by three conditions: (I.) volume growth, $\exists \;\gamma\in\R^+\;|\;\omega(Q_n^\star Q_n)< \gamma n^\di\;\forall\;n>0$; (II.) existence of the limit action $\lim_{n\to\infty} \omega(Q_n^\star A)$ for all $A\in\obs$; and (III.) invariance of the limit action under translations of $A$ in an appropriate uniform fashion (see Definition \ref{defiquasi}). If both $Q_n$ and $Q^\star_n$ form pseudolocal sequences, we call the linear functional $\widehat Q_\omega$, which acts on local observables as
\beq
	\widehat Q_\omega(A) = \frc12 \lim_{n\to\infty}
	\omega(\{Q^\star_n, A\}),
\eeq
a pseudolocal charge.

Note that we separate the notion of pseudolocal sequence from that of pseudolocal charge; the former represents the taking of the thermodynamic limit of actual operators, while the latter represents its action as a linear functional on a space of observables. In this way we avoid the problem of defining pseudolocal charges as operators in the thermodynamic limit.

Theorem \ref{propotwosided} states that pseudolocal charges are in bijection with the Hilbert space $\obsdh_\omega$, in such a way that for every $\widehat Q_\omega$ there exists a $A\in\obsdh_\omega$ with the property that
\beq
	\widehat Q_\omega(B) = \dbra A,B\dket_\omega
\eeq
for all $B$ (which now can be taken in $\obsdh_\omega$). The interpretation of this bijection is that the elements $A\in\obsdh_\omega$ are {\em densities} of the pseudolocal charges $\widehat Q_\omega$. This is clear in the ``local" case: indeed any sequence $Q$ formed by the finite partial sums of the series $\sum_{x} A_x$, for $A\in\obs$, is pseudolocal, and the associated pseudolocal charge acts as $\widehat Q_\omega(B) = \dbra A,B\dket_\omega$. Hence we find that the Hilbert space based on the sesquilinear form of susceptibilities is exactly the space of densities of pseudolocal charges.

There are similar results without symmetrizing: we may define left pseudolocal charges $Q_\omega(A) = \lim_{n\to\infty} \omega(Q_n^\star A)$ and right pseudolocal charges $Q_\omega^\star (A) = \overline{Q_\omega(A^\star)} = \lim_{n\to\infty} \omega(AQ_n)$, and similar statements hold for their relations with the Hilbert spaces based on sesquilinear forms of non-symmetrized susceptibilities (Theorems \ref{propoleft} and \ref{proporight}).

\begin{rema} Formally, $\dbra A,B\dket_\omega$ is the linear response of the average of $B$ to a small change of the potential associated to the ``charge'' $\sum_x A_x^\star$,
\beq\label{ovresp2}
	\dbra A,B\dket_\omega
	=
	\frc{\dd}{\dd \mu}\lt(\frc{
	\omega(e^{\frc\mu2 \sum_x A_x^\star}\, B\,
	e^{\frc\mu2 \sum_x A_x^\star})}{\omega(
	e^{\mu \sum_x A_x^\star})}\rt)_{\mu=0}.
\eeq
Note that symmetrization guarantees that the modified state still has nice properties under conjugations if $A^\star=A$.
\end{rema}

\subsection{Clustering for pseudolocal charges}

If $A,B,C\in\obs$, it is intuitively clear that when $B$ and $C$ (or more precisely their supports) are separated from each other by a large distance, then
\beq\label{ovclusform}
	\dbra A,BC\dket_\omega \approx
	\dbra A,B\dket_\omega\, \omega(C) +
	\dbra A,C\dket_\omega\, \omega(B).
\eeq
Indeed, the sum over $x$ of $\omega(A_x^\star BC) - \omega(A^\star)\omega(BC)$ should have nontrivial contributions only when $A_x$ lies in the region around $B$, or when $A_x$ lies around $C$, for in other regions clustering would apply and the contribution to the sum would be negligible. This is the case, and can be made precise (Theorem \ref{propocorr}). However, if $A$ is in the Hilbert space $\obsdh_\omega$, because a completion has been done, it does not have, generically, a finite support. Therefore, this na\"ive argument breaks down. Nevertheless, a certain weak property of clustering such as \eqref{ovclusform} does indeed hold (Definition \ref{defifunct}, Theorem \ref{propoclusform}), which guarantees in particular that
\beq
	\lim_{\dist(A,B)\to\infty} \widehat Q_\omega (AB)
	=\widehat Q_\omega (A)\omega(B)
	+ \widehat Q_\omega (B)\omega(A).
\eeq
This can be seen as an asymptotic derivation property.

\subsection{Pseudolocal states and Gibbs states}

We define a pseudolocal flow (Definition \ref{defiflow}) as follows: it is a family of uniformly bounded clustering states $\{\omega_s:s\in[0,1]\}$, with the property that there exists some uniformly bounded family of pseudolocal charges $\{\widehat Q_{s}:s\in[0,1]\}$ with respect to $\omega_s$ such that
\beq\label{ovflow}
	\omega_{s_+}(A)-\omega_{s_-}(A) = \int_{s_-}^{s_+}
	\dd s\,\widehat Q_{s}(A)\quad
	\forall\;0\leq s_-<s_+\leq 1
\eeq
for every local observable $A$; a Lebesgue integral is used on the right-hand side. We may also ask for clustering to hold uniformly  (see Definition \ref{defiflow}), or for the function $s\mapsto \omega_s(A)$ to be analytic. In the latter case, we have
\beq
	\frc{\dd}{\dd s} \omega_s(A) = \widehat Q_{s}(A)\quad
	\forall\;s.
\eeq
In this case, for every local observable $A$ there is a differentiable path whose tangent is described by the action of pseudolocal charges on $A$, which justifies the interpretation as a ``flow'' (on some manifold of clustering states).

\begin{figure}
\bc\includegraphics[height=3cm]{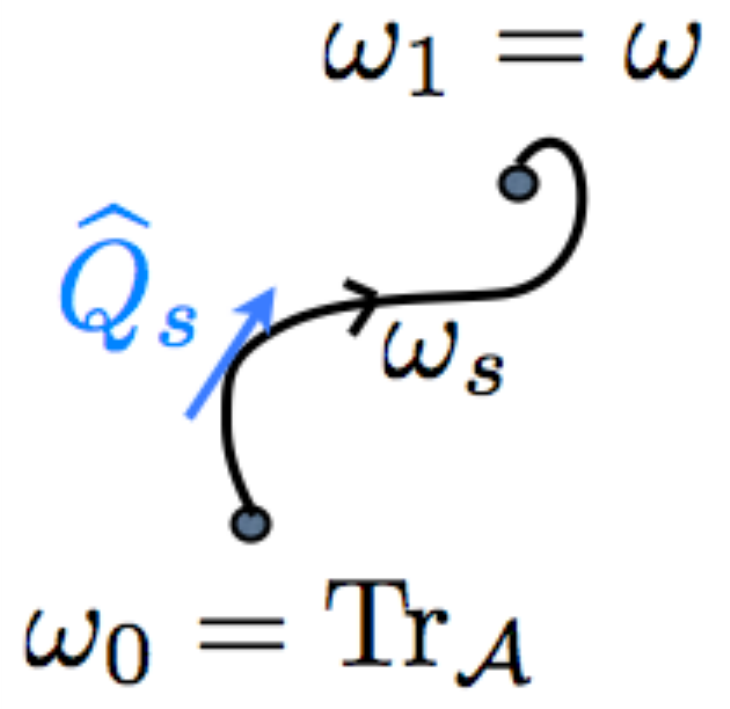}\ec 
\caption{A graphical representation of a pseudolocal state $\omega$.}
\label{fig}
\end{figure}

A pseudolocal state (Definition \ref{defistates}) is then defined as a state $\omega$ which is connected to the infinite-temperature state $\Tr_\obsc$ by an appropriately uniform pseudolocal flow $\{\omega\}$; here ``connected'' means that $\omega_0 = \Tr_\obsc$ and $\omega_1 = \omega$. See Figure \ref{fig}. We show that any thermal Gibbs state associated to a local Hamiltonian is a pseudolocal state (Theorem \ref{theothermal}), for all large enough temperatures, in particular temperatures above a universal lower bound $\beta_*^{-1}$ \cite{KGKRE} if the dimension $\di$ is larger than 1.

Pseudolocal states have very nice properties under finite time evolutions, which Gibbs states generically do not. Consider a dynamics $\tau_t:\obsc\to\obsc$ (for $t\in\R$) generated by a local Hamiltonian. It turns out that $\tau_t(A)$ makes sense as an element of the Hilbert space $\obsdh_{\omega}$ for any local observable $A$ and any time $t$ (Theorem \ref{propotime}). Further, if $\omega$ is a pseudolocal state, then $\omega\circ\tau_t$ also is, for every $t\in\R$ (Theorem \ref{theowb}): the family of pseudolocal states is preserved under time evolution.

\begin{rema} \label{remapexp} A non-rigorous interpretation of a pseudolocal state is as follows: let $q_s\in\obsdh_{\omega_s}$ be the density of $\widehat Q_s$, and $q_{s,x}$ its translate by $x\in\Z$. Then, formally, the pseudolocal state arises from a modification of the state $\Tr_\obsc$ by flows of operators controlled by the ``charge'' $\sum_x q_{s,x}^\star$,
\beq\label{ovint1}
	\omega_s(A) = \frc{\Tr_\obsc\lt(U_s\, A \,V_s\rt)}{
	\Tr_\obsc \lt(U_s\, V_s\rt)}
\eeq
with
\beq\label{ovint2}
	\frc{\dd}{\dd s} U_s = \frc12 U_s \,\sum_x q_{s,x}^\star,\quad
	\frc{\dd}{\dd s} V_s = \frc12 \sum_x q_{s,x}^\star\, V_s.
\eeq
The solutions to these may be expressed as path-ordered exponentials, for instance
\[
	U_s =\overrightarrow{\Pexp}\lt[\frc12\int_0^s \dd u\,\sum_x q_{u,x}^\star\rt],\quad
	V_s=\overleftarrow{\Pexp}\lt[\frc12\int_0^s \dd u\,\sum_x q_{u,x}^\star\rt].
\]
This is a generalization of the exponential on the right-hand side in \eqref{ovresp2} to families of generically non-commuting charges. Note that by the cyclic property of the trace, $\omega_s$ does take, for any $s$, the usual form with density matrix $V_sU_s$.
\end{rema}

\subsection{Generalized Gibbs ensembles}

Finally, we propose a rigorous definition of generalized Gibbs ensembles (GGEs) (Definition \ref{defiGGE}): a GGE with respect to a Hamiltonian $H=\sum_x h_x$ (seen as the sequence of its partial sums), where $h=h^\star$ is a local observable, is a pseudolocal state $\omega$, whose associated flow $\{\omega_s:s\in[0,1]\}$ and pseudolocal charges $\{\widehat Q_s:s\in[0,1]\}$ are almost everywhere preserved by the $H$-dynamics: for almost all $s\in[0,1]$,
\beq
	\omega_s(\rep H(A))=0,\quad \widehat Q_s(\rep H(A)) = 0
	\quad \forall\;A\in\obs
\eeq
where $\rep H(A)$ is the commutator $[H,A]$ (this is a local observable, so the above makes sense). Clearly, any thermal Gibbs state with respect to $H$ is, at high enough temperatures, a GGE with respect to $H$. But also, if $H$ is {\em completely mixing}, in the sense that its only conserved pseudolocal charges are those generated by multiples of $H$ itself (Definition \ref{defimixing}), then a GGE with respect to $H$, with appropriate analytic properties, is a thermal Gibbs state with respect to $H$ (Theorem \ref{theoGGEgibbs}). This is very similar in spirit to the eigenstate thermalization hypothesis.

Consider a pseudolocal state $\omega$ and a dynamics $\tau_t$, associated to a Hamiltonian $H$ with local density as above. Assume that the clustering property holds uniformly on the flow and for all large enough times. Assume also that $\lim_{t\to\infty}\dbra \tau_t(A),B\dket_{\omega_s}$ exists for all local observables $A,B$ and almost all $s$; that is, {\em all dynamical susceptibilities on the flow have a large-time limit}. In this case, the limit state $\lim_{t\to\infty}\omega\circ \tau_t$ exists ($\star$-weakly) and is a (weak) GGE with respect to $H$ (Theorem \ref{theoGGE}).

Combining all this, we also obtain a re-thermalization result: if the long-time evolution of a thermal Gibbs state with respect to some Hamiltonian $G$, evolved with some unrelated Hamiltonian $H$ that is completely mixing, satisfies uniform clustering properties, if the long-time dynamical susceptibilities exist, and if the long-limit of the state satisfies appropriate analyticity conditions, then the long-time limit is a thermal Gibbs state with respect to $H$ (Corollary \ref{theore}).

\begin{rema}\label{remaGGEov}
From Remark \ref{remapexp}, a GGE has the non-rigorous interpretation of a state obtained by path-ordered exponentials of charges that are all conserved by the dynamics. If the charges are, in some sense, in involution, the path-ordered exponentials $U_s$ and $V_s$ become ordinary exponentials. Using the cyclic property of the trace, and expressing all conserved charges $Q_s$ in terms of a countable basis, one then recovers, formally, the usual expression of GGEs (see also Remark \ref{remaGGE}).
\end{rema}

\section{Observables and states}\label{sectalg}

The basis for many rigorous formulations of quantum statistical mechanics are quasi-local $C^\star$ algebras, seen as algebras of observables, and positive linear functionals on these, seen as the states \cite[Sect. 2.6]{BR1}. Below, $\obsc$ is such an algebra of observables, and $\obs$ is the subspace of local observables. Let us further specify certain aspects of this algebra, and recall the main ingredients.

Let $\di\in\N$ and $\Z^\di$ be the $\di$-dimensional cubic lattice. Let $\obs$ be a normed linear space over $\C$, the norm being denoted $||\cdot||$, with a particular element $\1\in\obs$ of norm 1. We shall think of $\obs$ as the space of {\em local observables}. For every finite subset $X\subset \Z^\di$ let $\obs_X\subset \obs$ be a finite-dimensional subspace of $\obs$, with the properties that $\obs_{\emptyset} = \C\1$, that $X\subset Y\Rightarrow \obs_X\subset \obs_Y$, that $\obs_X \cap \obs_Y = \obs_{X\cap Y}$, and that the inductive limit is $\lim_{X\to \Z^\di} \obs_X=\obs$. From these properties, we see that for every $A\in\Or$ there exists a unique minimal subset $X\subset\Z^\di$ such that $A\in\obs_X$ (that is, if $A\in\obs_Y$ then $Y\supset X$); this will be referred to as the support of $A$, denoted $X=\supp(A)$.

Further, let the Cauchy completion $\obsc$ of $\obs$ be a unital $C^\star$-algebra where $\1$ is the unit, with the properties that every $\obs_X$ is a $C^\star$-subalgebra of $\obsc$, and that if $X,Y\subset \Z^\di$ are disjoint finite subsets, then $AB = BA$ and $\supp(AB)=\supp(A)\cup \supp(B)$ for all $A\in\obs_X$ and $B\in\obs_Y$. Note that in general $\supp(AB) \subset \supp(A)\cup \supp(B)$, and $\supp(A^\star) = \supp(A)$.

We assume that the structure is invariant under translation. That is, for every $x\in\Z^\di$ there exists a $\star$-automorphism $\iso_x:\obsc\to\obsc$, with the properties that $\supp(\iso_x(A)) = \supp(A)+x\;\forall\, x\in\Z^\di,\,A\in\obs$  (where $\supp(A)+x := \{y+x:y\in \supp(A)\}$), and such that $\iso_x\circ \iso_y = \iso_{x+y}\;\forall\;x,y\in\Z^\di$.

A state $\omega$ on the unital $C^\star$-algebra $\obsc$ is, as usual, a positive linear functional $\omega:\obsc\to\C$ such that $\omega(\1)= 1$. In particular, it is continuous and bounded, with
\beq\label{boundomega}
	|\omega(A)|\leq ||A||\quad\mbox{and}
	\quad \overline{\omega(A)} = \omega(A^\star)
	\quad \forall\; A\in\obsc
\eeq
(where $\overline{\omega(A)}$ is the complex conjugate of $\omega(A)$). A translation-invariant state is a state that is invariant under $\iso_x$; that is, $\omega(\iso_x(A)) = \omega(A)$ for all $A\in\obsc$ and $x\in\Z^\di$. 

Note that an example of $\obs$ where all above properties are satisfied is that of the inductive limit $\bigotimes_{n\in\Z^\di}\obs_{\{n\}}$ of the tensor product of copies $\obs_{\{n\}}$ of a finite-dimensional space of square matrices with the usual norm, see for instance \cite{BR1}. We will specialize to this case in Section \ref{secttime}.

Finally, it is convenient to have various simple geometric notions. The distance between $x$ and $y$ on the hypercubic lattice $\Z^\di$ is the minimal number of edges over all paths connecting $x$ with $y$, given by $\dist(x,y)= \sum_{i=1}^\di |x_i-y_i|$ (here and below we denote the components of $x\in\Z^\di$ as $x_i,\,i\in\{1,\ldots,D\}$). We define the size of an observable $A\in\obs$ as the cardinal of its support, $\siz A:=|\supp(A)|$, the distance between $A,B\in\obs$ as the distance between their supports, $\dist(A,B):=\dist(\supp(A),\supp(B)) = \min\{\dist(x,y):x\in\supp(A),\,y\in\supp(B)\}$, and the diameter of $A\in\obs$ as the maximal distance between points in its support, $\diam(A):=\diam(\supp(A))=\max\{\dist(x,y):x,y\in\supp(A)\}$. We will adopt the convention that distance $\dist(A,\1)=\infty$, representing the fact that the element $\1$ is supported on any subset of $\Z^\di$. The ball of radius $d$ centered at 0 will be denoted by $\ball(d) = \{x\in\Z^\di:\dist(x,0)\leq d\}$. Its area will be denoted by $\are_\di(d) = |\{x\in\Z^\di:\dist(x,0)= d\}|$. This is a polynomial of degree $\di-1$ that is positive for all $d>0$, with for instance $\are_1(d) = 2$, $\are_2(d)=4d$ and $\are_3(d)=2+4d^2$.

\section{Clustering} \label{sectclust}

The property of clustering represents the fact that any two elements of $\obs$ become less correlated as their distance increases. It has been widely studied. A weak form of clustering, ergodicity, is true for every translation-invariant extremal state, and a stronger form, mixing, is true for every extremal Gibbs (KMS) state (see the nice exposition given in \cite[Chap IV]{Israel}). Stronger still, exponential clustering -- and related analyticity properties -- of the unique Gibbs state with respect to a local Hamiltonian in quantum chains was first established by Araki \cite{Araki69}. In dimensions higher than one, exponential clustering and analyticity holds at high enough temperatures, see the books \cite{Ruelle,BR1,Simon} and references therein. Later studies in quantum systems include \cite{GN98,Mat02}, and a recent characterization that will be of immediate use for us is given in \cite{KGKRE}. Exponential clustering was also proved in gapped quantum chains \cite{Hastings04,NS05,Hast06}, see also \cite{Moh15,Na}.

We define a general notion of clustering which contains, but is weaker than, the exponential clustering that is generic in high-temperature Gibbs states and gapped ground states.

\begin{defi}\label{deficlust} {\em (Clustering states)} Let $\omega$ be a translation-invariant state on $\obsc$ with the following additional property: there exists $\nu:\N\to\R^+$ and $f:\Z\to\R^+$ such that for every $\ell\in\N$ and every $A,B\in\obs$ with $\siz{A},\siz{B}<\ell$, the following inequality holds:
\beq\label{clusterstate}
	\big|\omega(AB) - \omega(A)\omega(B)\big|
	\leq \nu(\ell)\,||A|| \,||B||\,f(\dist(A,B)).
\eeq
We say that $\omega$ is algebraically clustering with power $p>\di$ (algebraically $p$-clustering) if $f(d)\leq d^{-p}$ for all $d>0$, that it is exponentially clustering with exponent $\alpha>0$ (exponentially $\alpha$-clustering) if $f(d) \leq e^{-\alpha d}$ for all $d>0$, and that it is clustering if it is algebraically $p$-clustering for some $p>\di$.
\end{defi}
Choosing $\omega$ to be a normalized ($\omega(\1) = 1$), translation-invariant linear functional on $\obs$ -- that is, omitting the requirements of positivity and boundedness -- the above defines a {\em clustering linear functional} on $\obs$.

Obviously, $\nu(\ell)$ can be chosen to be non-decreasing. The sum of $\are_\di(d)f(d)$ will be denoted $F$:
\beq
	F:=\sum_{d=0}^\infty \are_\di(d)f(d) <\infty.
\eeq
By a normalization of $\nu(\ell)$ we may assume $F=1$, but it will be simpler to keep $F$ arbitrary.

Note that the clustering property applies solely to local observables, because sizes need to be finite. In this and the next section, we will discuss only local observables $\obs$ and their completion according to norms that are different from that of $\obsc$.

Definition \ref{deficlust} means that the quantity $\omega(AB) - \omega(A)\omega(B)$ decays fast enough with the distance between $A$ and $B$, in a way that is uniform over all observables of norms and sizes with fixed upper bounds. It also implies the weaker statement of the uniform existence of the limit $\lim_{\dist(A,B)\to\infty}\omega(AB) = \omega(A)\omega(B)$. If $B=\1$, then $\omega(AB) - \omega(A)\omega(B)$ vanishes, and the inequality is in agreement with the convention according to which the distance $\dist(A,\1)=\infty$.

When discussing time evolution, it will be important to have slightly stronger conditions on clustering: we will need to control the function  $\nu(\ell)$, making sure that it can be chosen to grow algebraically with $\ell$ (no matter how big the power). Clustering states with such control on $\nu(\ell)$ will be referred to as sizably clustering.
\begin{defi}\label{defisiz} {\em (Sizably clustering states)} Let $\omega$ be a clustering state with decay function $f$. We say that $\omega$ is sizably clustering if there exist $\nu,a>0$ such that for every $\ell>0$ and every $A,B\in\obs$ with $\siz{A},\siz{B}<\ell$, we have
\beq\label{sizclusterstate}
	\big|\omega(AB) - \omega(A)\omega(B)\big|
	\leq \nu\ell^a\,||A|| \,||B||\,f(\dist(A,B)).
\eeq
\end{defi}
Again, omitting the requirements of positivity and boundedness, the above defines a {\em sizably clustering linear functional} on $\obs$.

When discussing families of states $\{\omega\}$, we will say that a family is uniformly (sizably) clustering if there is a single $\nu(\ell)$ (a single pair $\nu,a$) and a single $p>\di$ applicable to all states in the family.

It is important to remark that any thermal state is a sizably exponentially clustering state, this holding for high enough temperatures if $\di>1$ (see Theorem \ref{theothermal} and the discussion around it). The same holds for ground states of gapped quantum chains, by theorems of Hastings, and of Nachtergaele and Sims \cite{Hastings04,NS05}.

Given a clustering linear functional, we may define a sesquilinear form $\bra \cdot,\cdot\ket_\omega: \obs\times\obs \to \C$ as follows:
\beq\label{form}
	\bra A,B\ket_\omega := \sum_{x\in\Z^\di}
	\big(\omega(\iso_x(A)^\star B) - \omega(A^\star)\omega(B)\big).
\eeq
The series is convergent, and for $\omega$ a clustering state, the resulting form is positive semi-definite.

\begin{theorem} \label{propoconv} Let $\omega$ be a clustering linear functional. For every $\ell,w>0$, the series in \eqref{form} converges absolutely and uniformly on $\{A,B\in\obs:|A|,|B|<\ell,\,||A||,||B||<w\}$. For every $\ell>0$ and every $A,B\in\obs$ with $|A|,|B|<\ell$,
\beq
	|\bra A,B\ket_\omega| \leq F\nu(\ell)\,\siz{A}\,\siz{B}\, ||A||\,||B||.
\eeq
\end{theorem}
\proof
This holds as a consequence of the clustering property \eqref{clusterstate}. It can be established by noting that given $A$ and $B$, for every value of $d\in\N$, there is at most $\are_\di(d)|A|\,|B|$ values of $x$ solving the relation $\dist(A_x,B)=d$. These values of $x$ correspond to the $\siz{A}\,\siz{B}$ pairs of points in $\supp(\iso_x(A))\times \supp(B)$ which can determine the minimal distance between the sets $\supp(\iso_x(A))$ and $\supp(B)$ (and given a pair and a distance, the point belonging to $\supp(\iso_x(A))$ can be anywhere on the sphere centered at the point belonging to $\supp(B)$, whence the factor $\are_\di(d)$). Thus the sum over $x\in\Z^\di$ of the absolute values of the summand in \eqref{form} can be bounded by a sum over non-negative integers $d$ of $\are_\di(d)\nu(\ell) f(d)\siz{A}\,\siz{B}\,||A||\,||B||$.
\eproof

\begin{lemma}\label{lemsum}
Let $\omega$ be a clustering linear functional, let $A\in\obs$ with $\omega(A)=0$, and let $B_L:=\sum_{x\in [1,L]^\di} \iso_x(A)$ for all $L\in\N$. Then,
\beq
	\lim_{L\to\infty} L^{-\di} \omega(B_L^\star B_L) = \bra A,A\ket_\omega.
\eeq
\end{lemma}
\proof By translation invariance
\[
	\omega(B_L^\star B_L) = \sum_{x\in[-L,L]^\di}\lt(\prod_{i=1}^\di(L-|x_i|)\rt) \omega(\iso_x(A)^\star A).
\]
In general, given $g(x)\in\C$ and $p>\di,\,\nu>0$ such that $|g(x)|\leq\nu \dist(x,0)^{-p}$ for all $x\in\Z^\di$, we have
\beqa
	\lefteqn{\lt|L^\di\sum_{x\in\Z^\di} g(x)-
	\sum_{x\in[-L,L]^\di}\lt(\prod_{i=1}^\di(L-|x_i|)\rt) g(x)\rt|} \hspace{3cm}&&\n
	&\leq&
	L^{\di-1} \sum_{x\in[-L,L]^\di} \sum_{i=1}^\di |x_i|\,|g(x)|
	+ L^\di\sum_{x\in\Z^\di\setminus[-L,L]^\di} |g(x)| \n
	&\leq& L^{\di-1}\sum_{x\in\ball(\di L)} \sum_{i=1}^\di |x_i| |g(x)| 
	+\nu\,L^\di \sum_{x\in\Z^\di\setminus \ball(L)} |g(x)| \n
	&\leq& \nu L^{\di-1}\sum_{d=0}^{\di L} \are_\di(d) d^{-p+1}
	+\nu\,L^\di \sum_{d=L}^\infty \are_\di(d) d^{-p}. \no
\eeqa
In the first step we used $0\leq L^\di - \prod_i (L-|x_i|) \leq L^{\di-1}\sum_i |x_i|$. In the second step, we used the fact that $\ball(L)\subset [-L,L]^\di\subset\ball(\di L)$ (for the latter inclusion: corners of the cube are a distance $\di L$ from its center). Since $\are_\di(d)$ is a polynomial of degree $\di-1$, the first sum is bounded by $aL^{\di-1}+b L^{2\di-p}$ for some $a>0,\,b>0$ (the first term dominates for $p>\di+1$, the second for $p\in(\di,\di+1]$), and the second sum is bounded by $c L^{2\di-p}$ for some $c>0$ (for all $p>\di$). As a consequence, multiplying by $L^{-\di}$, the limit $L\to\infty$ is zero, and we find
\beq\label{pr15}
	\sum_{x\in\Z^\di} g(x)= \lim_{L\to\infty} L^{-\di}
	\sum_{x\in[-L,L]^\di}\lt(\prod_{i=1}^\di(L-|x_i|)\rt) g(x).
\eeq
Because of the clustering property \eqref{clusterstate}, and bounding by $\siz{A}^2$ the number of pairs of points in $\supp(\iso_x(A))\times\supp(A)$ that may determine the distance $\dist(\iso_x(A),A)$ (as per the argument used in Theorem \ref{propoconv}), we may apply the above to $g(x) = \omega(\iso_x(A)^\star A)$. This shows the lemma. \eproof

\begin{theorem} \label{propopos} Let $\omega$ be a clustering state. The form $\bra \cdot,\cdot\ket_\omega$ is positive semi-definite: $\bra A,A\ket_\omega\geq0$ for all $A\in\obs$.
\end{theorem}
\proof Without loss of generality (by considering $A-\omega(A)\1$ instead of $A$) we assume $\omega(A)=0$. Let $B_L$ be as in Lemma \ref{lemsum}. Positivity $\omega(B_L^\star B_L)\geq 0$ then implies the proposition.
\eproof

By translation invariance of the state $\omega$, the form  \eqref{form} is translation invariant,
\beq
	\bra \iso_x(A),B\ket_\omega = \bra A,\iso_x(B)\ket_\omega = \bra A,B\ket_\omega.
\eeq
Also, it is degenerate, as $\bra \1,A\ket_\omega = 0$ and $\bra \iso_x(B)-B,A\ket=0$ for all $A,B\in\obs$. In general the form is non-degenerate on a quotient space $\obsn_\omega=\obs/ \obsd_\omega$ where $\C\1\oplus \{\iso_x(B)-B:x\in\Z^\di,\,B\in\obs\}\subset \obsd_\omega$. The induced inner product on the quotient space $\obsn_\omega$ gives rise to a Hilbert space $\obsh_\omega$ by Cauchy completion of $\obsn_\omega$ in the induced norm. We will denote by $[\cdot]_\omega:\obs\to\obsn_\omega$ the coset map, but for simplicity we will use the same notation $\bra A,B\ket_\omega$ for $A,B$ in $\obs$ or in $\obsh_\omega$. We will understand $\obsn_\omega$ as the subspace of ``local elements'' of the Hilbert space $\obsh_\omega$.

The form $\bra \cdot,\cdot\ket_\omega$, built from the clustering state $\omega$, itself has a clustering property. Although this property does not play any role in the derivation of our main results, it is useful in order to characterize the pseudolocal charges themselves, in particular their own clustering property. Here we provide a general definition of clustering forms, independent of the state $\omega$, and then establish their properties and the relation between clustering states and forms. We defer the proofs to Appendix \ref{appform}.

\begin{defi}\label{defiform} {\em (Clustering forms)} Let $\omega$ be a state on $\obs$ and $\bra\cdot,\cdot\ket$ be a translation-invariant positive semi-definite sesquilinear form on $\obs$ with the following additional properties:
\bi
\item[\rm I.] There exist $\nu:\N\to\R^+$ and $f:\Z\to\R^+$ such that for every $\ell\in\N$ and every $A,B,C\in\obs$ with $\siz{A},\siz{B},\siz{C}<\ell$, we have
\beq\label{clusterform}
	\big|\bra A,BC\ket - \bra A,B\ket\omega(C) - \bra A,C\ket \omega(B)\big| \leq \nu(\ell)\,
	||A|| \,||B|| \,||C||\,f(\dist(B,C)-\diam(A)).
\eeq
\item[\rm II.] For every $\ell>0$, there exists $\kappa>0$ such that for every $A,B\in\obs$ with $|A|,|B|<\ell$,
\beq
	|\bra A,B\ket_\omega| \leq \kappa\, ||A||\,||B||.
\eeq
\ei
We say that $\bra\cdot,\cdot\ket$ is a (algebraically, exponentially) clustering form on $\Or$ with respect to the state $\omega$ if $\nu(\ell)$ and $f(d)$ are as specified in Definition \ref{deficlust}.
\end{defi}
This implies in particular that for every $A\in \obs$ the limit  $\lim_{\dist(B,C)\to\infty} \bra A,BC\ket =\bra A,B\ket\omega(C) + \bra A,C\ket \omega(B)$ exists uniformly over all observables $B,C$ of norms and sizes with fixed upper bounds. Setting $C=\1$, the convention that $\dist(B,\1)=\infty$ means that, implied in the definition of a clustering form, there is the equality $\bra A,\1\ket=0$ for all $A\in\obs$.

Let $\obsd$ be the degenerate subspace of $\bra\cdot,\cdot\ket$, let $\obsn$ be the quotient space $\obs/\obsd$ on which $\bra\cdot,\cdot\ket$ is an inner product, and let $\obsh$ be its Cauchy completion. We will denote the norm by $||A||_{\obsh}:=\sqrt{\bra A,A\ket}$. We may wonder if a clustering property holds for all $A$ in the Hilbert space $\obsh$. Because uniformity of the inequality in \eqref{clusterform} holds only for $A$ of fixed maximal size, we cannot directly apply the inequality to generic elements of the Hilbert space. However, it turns out that a weaker clustering statement holds on $\obsh$, which we express via what we will refer to as weakly clustering linear functionals.

\begin{defi}\label{defifunct} {\em (Weakly clustering linear functionals)} A weakly clustering linear functional with respect to $\omega$ is a translation-invariant linear functional $P$ on $\obs$ with $P(\1)=0$, satisfying the following additional properties:
\bi
\item[\rm I.] For every $\ell>0$ and $w>0$, the following limit exists uniformly on $A,B\in\obs$ with $\siz{A},\siz{B}<\ell$ and $||A||,||B||<w$,
\beq\label{clusterweak}
	\lim_{\dist(A,B)\to\infty} P(AB) =
	P(A)\omega(B) + P(B) \omega(A).
\eeq
\item[\rm II.] For every $A,B\in\obs$, the following limit exists and gives zero:
\beq\label{clusterhalf}
	\lim_{L\to\infty} 
	L^{-\di+\frc12}\sum_{x\in\ball(L)}\Big(P(\iso_x(A)B)
	-P(A)\omega(B) - P(B) \omega(A)\Big) = 0
\eeq
\ei
\end{defi}

\begin{theorem}\label{propoclusform}
Let $\bra\cdot,\cdot\ket$ be a clustering form on $\obs$ with respect to $\omega$, and let $\obsh$ be the associated Hilbert space (the Cauchy completion of the inner-product quotient space). Then for every $A\in\obsh$ the functional $\bra A,\cdot\ket$ is a weakly clustering linear functional on $\obs$.
\end{theorem}

Also, the natural form built out of a clustering state is a clustering form, albeit with a weaker clustering type.

\begin{theorem} \label{propocorr} Let $\omega$ be an algebraically $p$-clustering state with $p>2\di$ (exponentially $\alpha$-clustering state with $\alpha>0$) on $\obs$. Then $\bra\cdot,\cdot\ket_\omega$ is an algebraically $q$-clustering form (exponentially $\beta$-clustering form) on $\obs$ with respect to $\omega$, for all $\di<q<p-\di$ ($0<\beta<\alpha/2$).
\end{theorem}

See Appendix \ref{appform} for the proofs of these theorems.

\section{Pseudolocal charges} \label{sectquasi}

In the spirit of developing parts of a general theory of pseudolocal charges, in this section we consider three types of pseudolocal charges: left, right and two-sided. Even though only the latter type will be used when discussing time evolution and the GGE in the next section, it is instructive to introduce it via the former ones, displaying the full relationships between them that involves the $\star$-involution.

\subsection{Left pseudolocal charges} \label{ssectquasileft}

\begin{defi}\label{defiquasi}
Let $\omega$ be a clustering state on $\obs$, and let $Q=\{n\in\N \mapsto Q_n\in \obs\}$ be a sequence of observables such that $Q_n\in\obs_{\ball(n)}$ for all $n$. We say that such a sequence is  pseudolocal with respect to $\omega$ if the zero-average sequence $\{n\mapsto Q_n-\omega(Q_n)\1\}$ is pseudolocal, and pseudolocality in the case $\omega(Q_n)=0\,\forall\, n$ is defined by the following conditions:
\bi
\item[\rm I.] {\em Volume growth.} There exists $\gamma>0$ such that
\beq\label{ql1}
	\omega(Q_n^\star Q_n) \leq \gamma n^\di
\eeq
for all $n>0$.
\item[\rm II.] {\em Limit action.} For every $A\in\obs$, the limit
\beq\label{ql2}
	Q_\omega(A):=\lim_{n\to\infty} \omega(Q_n^\star A)
\eeq
exists in $\C$.
\item[\rm III.] {\em Bulk homogeneity.} There exists $0<k<1$ such that for every $A\in\obs$,
\beq\label{ql3}
	\lim_{n\to\infty}
	\max_{x,y\in\ball(kn)} |\omega(Q_n^\star \iso_x(A))-\omega(Q_n^\star\iso_y(A))|
	=0.
\eeq
\ei
The limit action $Q_\omega$ will be referred to as a left pseudolocal charge with respect to $\omega$, and we will denote the set of these by $\obsq_\omega$.
\end{defi}

Note that $(\cdot)_\omega$, as a mapping from sequences $Q$ to elements $Q_\omega\in\obsq_\omega$, is anti-linear. The set $\obsq_\omega$ forms a linear space: indeed conditions II and III are linear, and condition I extends to linear combinations by positivity of the state and the Cauchy-Schwartz inequality, $\omega(Q_n^\star Q_m) \leq \sqrt{\omega(Q_n^\star Q_n)\omega(Q_m^\star Q_m)}$. It is a space of functionals on $\obs$. Any sequence $Q$ giving rise to $Q_\omega$ will be referred to as a kernel of $Q_\omega$ ($Q$ is not unique).

In the next theorem we show that there is a bijection $\dens:\obsq_\omega\to\obsh_\omega$ whereby $Q_\omega(\cdot) = \bra \dens(Q_\omega),\cdot\ket_\omega$  (for simplicity, the dependence of $\dens$ on the state $\omega$ is kept implicit). We will refer to $\dens(Q_\omega)$ as the density of the left pseudolocal charge $Q_\omega$, and one can see that $\dens$ is anti-linear. This bijection implies that the linear action of $Q_\omega$ may be extended to $\obsh$, and allows us to put a norm on $\obsq_\omega$ induced via $\dens$ by that on $\obsh_\omega$. This is the same norm as that on $\obsq_\omega$ viewed as the dual $\obsh_\omega^*$:
\beq
	||Q_\omega||_{\obsq_\omega} =
	\sup_{A\in\obsh_\omega\setminus\{0\}} \lt(\frc{|Q_\omega(A)|}{
	||A||_{\obsh_\omega}}\rt)
	= \sup_{A\in\obsh_\omega\setminus\{0\}} \lt(
	\frc{|\bra \dens(Q_\omega),A\ket_\omega|}{
	||A||_{\obsh_\omega}}\rt)
	= ||\dens(Q_\omega)||_{\obsh_\omega}\label{normQ}
\eeq
where the last step is by the fact that the Cauchy-Schwartz inequality is saturated with $A = \dens(Q_\omega)$. If $\dens(Q_\omega)$ is in the subspace $\obsn_\omega$ of local elements, we say that $Q_\omega$ is a local charge, and we remark that local charges form a dense subspace of $\obsq_\omega$. 

Note that given $A\in\obs_{\ball(u)}$, $u\in\N$, the sequence $Q_n=\sum_{x\in\ball(n-u)} \iso_x(A)$ (for $n\geq u$), $Q_n=0$ (otherwise), gives rise to a local charge $Q_\omega$ with $\dens(Q_\omega)=[A]_\omega$ (the coset of $A$ with respect to $\obsd_\omega$), and that any local charge may be obtained in this way. We refer to such sequences as local sequences, denoted by
\beq\label{localseq}
	Q = \sum_{x\in\Z^\di} \iso_x(A)\qquad\mbox{(local sequence),}
\eeq
and to $A\in\obs$ as the density of the local sequence. The series in \eqref{localseq} is formal, as it is neither an element of the $C^\star$ algebra of observables nor of the Hilbert space $\obsh_\omega$. We understand it as the sequence of its partial sums over finite balls.

In order to show the main theorem relating left pseudolocal charges to the Hilbert space $\obsh_\omega$, we will make use of the following lemma:
\begin{lemma} \label{lemdelta} Let $\delta_{n,j}\geq 0,\, n,j\in\N$ be such that $\lim_{n\to\infty} \delta_{n,j}=0$ for all $j\in\N$, and let $j\mapsto u_j\in\N$ be some sequence of integers. Then there exists an unbounded increasing sequence $n\mapsto j_n\in\N$ such that $n>u_{j_n}$ for all $n$ large enough and $\lim_{n\to\infty}\delta_{n,j_n}=0$.
\end{lemma}
\proof The fact that $\lim_{n\to\infty} \delta_{n,j}=0$ means that for all $\delta>0$ and all $j\in\N$ there exists $n\in \N$ such that $\delta_{n',j}<\delta$ for all $n'\geq n$. Consider the sets $S(n,\delta):=\{j:\delta_{n',j}<\delta\;\forall\;n'\geq n\}\cap\{j:n>u_j\}$ for all $n\in\N$, $\delta>0$ ($S(n,\delta)$ may be empty). Clearly $S(n,\delta)\subset S(n+1,\delta)$ for all $n,\delta$. Let $J(n,\delta):=\sup(\{1\}\cup S(n,\delta))$ (which may be infinite). For any $\delta>0$, this is increasing in $n$ and satisfies $\lim_{n\to\infty}J(n,\delta)=\infty$.

If $\forall \;\delta>0\;\exists\; N=N(\delta)\in\N\;|\;J(N,\delta)=\infty$, then we can define $\delta_n:=\min\{\delta>0:n\geq N(\delta)\}$, which is decreasing in $n$ and satisfies $\lim_{n\to\infty} \delta_n=0$, and we can choose any $j_n\in S(n,\delta_n)$ to be strictly increasing, and the lemma is proved.

Let us then assume that $J(n,\delta)<\infty$ for all $n\in\N$, $\delta>0$ small enough. By definition, we have $\delta_{n',J(n,\delta)}<\delta\;\forall\;n'\geq n$ and $n>u_{J(n,\delta)}$ for all $n$ such that $J(n,\delta)>1$ (this latter inequality guarantees that $S(n,\delta)$ is non-empty). Fix $\delta_0>0$ small enough and construct the strictly increasing sequence $n_m$ recursively by $n_1=1$ and $n_m = \min \{n:n> n_{m-1},\,J(n,\delta_0/m)> J(n_{m-1},\delta_0/(m-1))\}$ for all integers $m\geq 2$. Since $J(n,\delta)$ is increasing in $n$ and tends to $\infty$, the set on which the minimum is taken is non-empty and each $n_m$ is in $\N$. Let $j_m':=J(n_m,\delta_0/m)$. By construction this is strictly increasing in $m$, and since $J(n_m,\delta_0/m)>J(n_1,\delta_0)\geq 1$ for all $m\geq 2$, it is such that,  for all $m\geq 2$, we have $\delta_{n',j_m'}<\delta_0/m\;\forall\;n'\geq n_m$ and $n_m>u_{j_m'}$. Finally we construct the increasing sequence $m_n:=\max\{m:n_m\leq n\}\in\N$, noting that $\lim_{n\to\infty}m_n=\infty$, and we set $j_n = j_{m_n}'$. Then, for all $n$ large enough (specifically, all $n$ such that $m_n\geq 2$), we have $\delta_{n',j_n}<\delta_0/m_n\;\forall\;n'\geq n$ and $n>u_{j_n}$. This shows the Lemma. \eproof

\begin{theorem}\label{propoleft}
Let $\omega$ be a clustering state on $\obs$. There exists a bijection $\dens:\obsq_\omega\to\obsh_\omega$ such that, for every $Q_\omega\in\obsq_\omega$ and every $A\in\obs$,
\beq\label{Qomleft}
	Q_\omega(A) = \bra \dens(Q_\omega),A\ket_\omega.
\eeq
In particular, $Q_\omega$ can be extended to a continuous linear functional on $\obsh_\omega$.
\end{theorem}
\proof
$\obsq_\omega\to\obsh_\omega$. Let $Q=\{n\mapsto Q_n\}$ be a pseudolocal sequence, and $A\in\obs$ with $\omega(A)=0$. Let $k'=k/\di\in(0,1/\di)$. We have ($L_n := 2k'n+1$)
\beqa
	\omega(Q_n A) &=& \frc1{L_n^\di} \sum_{x\in[-k'n,k'n]^\di}
	\omega(Q_n^\star \iso_x(A)) + \delta_n \n
	&\leq& \sqrt{\frc{\omega(Q_n^\star Q_n)}{L_n^\di}}
	\sqrt{\frc1{L_n^\di}\sum_{x,y\in[-k'n,k'n]^\di}\omega(\iso_x(A^\star)\iso_y(A))} +\delta_n.\no
\eeqa
In the first line, $\delta_n$ is such that $\lim_{n\to\infty}\delta_n=0$ by homogeneity of the pseudolocal sequence: we note that the hypercube $[-k'n,k'n]^{\di}$ is contained into the ball $\ball(kn)$. The second line is obtained by the Cauchy-Schwartz inequality for the positive semi-definite sesquilinear form $(A,B) = \omega(A^\star B)$. Therefore, there exist $\gamma>0$ and $0<k<1$ such that for every $A\in\obs$ with $\omega(A)=0$,
\beq
	Q_\omega(A) \leq
	\sqrt{\frc{\gamma}{(2k')^\di}}\,
	|| A||_{\obsh_\omega}
\eeq
where we used Lemma \ref{lemsum}, the volume growth condition, and the existence of the limit action. Since $Q_\omega(\1)=0$, this holds for every $A\in\obs$. As a consequence, $Q_\omega$ is a bounded, hence continuous, linear functional on $\obs$ with respect to $||\cdot||_{\obsh_\omega}$, and it can be continued in a unique fashion to a continuous linear functional on $\obsh_\omega$. By the Reisz representation theorem, this defines an anti-linear map $\dens:\obsq_\omega\to\obsh_\omega$ which is injective.

$\obsh_\omega\to\obsq_\omega$. Let $A\in\obsh_\omega$ and consider a Cauchy sequence $A_j\in\obs,\,j\in\N$ with $\lim_{j\to\infty} [A_j]_\omega = A$ (such a sequence exists by the fact that $\obsn_\omega$ is dense in $\obsh$).  We may assume without loss of generality that $\omega(A_j)=0$. Denote by $u_j:=\di\diam(A_j)/2$ and let $x_j \in\Z^\di$ be such that $\supp(\iso_{x_j}(A_j))\subset [-u_j/\di,u_j/\di]^\di$. Let
\[
	S_{n,j}:=\lt[-\frc{n-u_j}{\di},\,\frc{n-u_j}{\di}\rt]^\di
\]
and consider
\beq\label{prQnj}
	Q_{n,j} := \sum_{x\in S_{n,j}} \iso_{x+x_j}(A_j)
\eeq
for all $j\in\N$ and $n>u_j$. It is such that $\supp(Q_{n,j})\subset[-n/\di,n/\di]^\di\subset\ball(n)$ (for all such $j,\,n$).

Fix some $B\in\obs$ and let us define
\beqa
	\delta_{n,j}' &=& \frc1{2n}\omega(Q_{n,j}^\star Q_{n,j}) - \bra A_j,A_j\ket_\omega\n
	\delta_{n,j}'' &=& \omega(Q_{n,j}^\star B) - \bra A_j,B\ket_\omega \n
	\delta_{n,j}'''&=& \max_{y,z\in\ball(\frc n{2\di})}
	|\omega(Q_{n,j}^\star \iso_y(B)) - \omega(Q_{n,j}^\star \iso_z(B))|.
\eeqa
By Lemma \ref{lemsum}, we have $\lim_{n\to\infty}\delta_{n,j}'=0$ for every $j$, and by Theorem \ref{propoconv}, we have $\lim_{n\to\infty}\delta_{n,j}''=0$ for every $j$. By translation invariance, we have
\[
	\omega(Q_{n,j}^\star \iso_y(B)) = \sum_{x\in S_{n,j}-y} \omega(\iso_{x+x_j}(A_j^\star )B),
\]
and with $T_{n,j}(y,z) := ((S_{n,j}-y) \cup (S_{n,j}-z)) \setminus ((S_{n,j}-y)) \cap (S_{n,j}-z))$ we find
\[
	\delta_{n,j}'''
	= \max_{y,z\in\ball(\frc n {2\di})}
	\Big|\sum_{x\in T_{n,j}(y,z)}\omega(\iso_{x+x_j}(A_j^\star )B)\Big|.
\]
Since the linear size of the hypercube $S_{n,j}$ grows linearly like $2n/\di$ as $n\to\infty$, and since $T_{n,j}(y,z)$ does not contain any point shared by $S_{n,j}-y$ and $S_{n,j}-z$, a ball of radius growing like $n/(2\di)$ is always omitted in $T_{n,j}(y,z)$, for any $y,z\in\ball(n/(2\di))$. That is, for any $j$, there exists a $q\in\N$ and $n_0\in\N$ such that
\[
	\ball\lt(\frc n{2\di}-q\rt)\;\bigcap\; \lt(\bigcup_{y,z\in\ball(\frc n{2\di})} T_{n,j}(y,z)\rt)=\emptyset
\]
for all $n>n_0$. Therefore, 
\[
	\delta_{n,j}'''\leq \sum_{x\in\Z^\di\setminus\ball(\frc n{2\di}-q)}
	|\omega(\iso_{x+x_j}(A_j^\star)B)|
\]
for all $n>n_0$. By the clustering property of the state we then have $\lim_{n\to\infty} \delta_{n,j}'''=0$ for every $j$.

Define $\delta_{n,j} = |\delta_{n,j}'| + |\delta_{n,j}''|+|\delta_{n,j}'''|$: it is such that $\lim_{n\to\infty}\delta_{n,j}=0$ for every $j$. Consider the sequence $n\mapsto j_n$ of Lemma \ref{lemdelta}. Then $\lim_{n\to\infty}\delta_{n,j_n}=0$, and so $\lim_{n\to\infty} \delta_{n,j_n}' = \lim_{n\to\infty} \delta_{n,j_n}''=\lim_{n\to\infty} \delta_{n,j_n}'''=0$; and $n>u_{j_n}$ for all $n\in\N$ large enough. Define
\beq\label{prQndef}
	Q_n:=Q_{n,j_{n}}.
\eeq
Then
\beqa
	\lim_{n\to\infty}
	\lt(
	\frc1{2n}\omega(Q_{n}^\star Q_{n}) -\bra A_{j_n},A_{j_n}\ket_\omega\rt)
	&=& 0 \n
	\lim_{n\to\infty}
	\lt(\omega(Q_{n}^\star B) - \bra A_{j_n},B\ket_\omega\rt) &=& 0\n
	\lim_{n\to\infty}
	\max_{y,z\in[-n/2,n/2]}
	|\omega(Q_{n}^\star \iso_y(B)) - \omega(Q_{n}^\star \iso_z(B))| &=&0.
\eeqa
The first implies that $Q$ satisfies the property of linear growth. The second implies that the limit action $Q_\omega$ exists and that
\beq\label{prqome}
	Q_\omega(B) = \bra A,B\ket_\omega
\eeq
for all $B\in\obs$, and therefore by continuous extension for all $B\in\obsh_\omega$. The third implies that $Q$ satisfies the property of homogeneity. Thus this gives a map $\obsh_\omega\to\obsq_\omega$: $A\mapsto Q_\omega$. This map is well defined as any two Cauchy sequences giving the same $A\in\obsh_\omega$ gives rise to the same linear functional $Q_\omega$ by \eqref{prqome}, it is injective, and its composition with $\dens$ gives the identity thanks to the first part of the proof. Hence it is $\dens^{-1}$, and we have shown that $\dens$ is a bijection.
\eproof

The Hilbert space structure also allows us to show that a pseudolocal charge can always be seen as the strongly converging limit of a sequence of local charges  (converging in the norm in $\obsh_\omega^*$), with the corresponding densities being similarly related (convergence in $\obsh_\omega$).

\begin{theorem}\label{propolocseqleft}
Let $j\mapsto A_j\in\obs$ be Cauchy sequence with respect to $||\cdot||_{\obsh_\omega}$ and $A\in\obsh_\omega$ the corresponding limit. Then the left pseudolocal charge $Q_\omega = \dens^{-1}(A)$ may be obtained as the strongly converging limit of the local charges built out of local sequences with densities $A_j$,
\beq
	Q_\omega = \lim_{j\to\infty}
	\lt(\sum_{x\in\Z^\di} \iso_x(A_j)\rt)_\omega.
\eeq
\end{theorem}
\proof Let $Q_j:=\dens^{-1}([A_j]_\omega)\in\obsq_\omega$ be the $j^{\rm th}$ local charge. Then, by \eqref{normQ},
\beqa
	||Q_j-Q_k||_{\obsq_\omega} = ||A_j-A_k||_{\obsh_\omega}
\eeqa
wherefore $j\mapsto Q_j$ forms a Cauchy sequence with respect to $||\cdot||_{\obsq_\omega}$. The limit is
\beq
	\lim_{j\to\infty} Q_j(B) = \lim_{j\to\infty}\bra A_j,B\ket_\omega
	=\bra A,B\ket_\omega = Q_\omega(B).
\eeq
\eproof

The following corollary, which characterizes the clustering properties of left pseudolocal charges, immediately follows from Theorems \ref{propoleft}, \ref{propoclusform} and \ref{propocorr}.

\begin{corol} \label{coroleft} A left pseudolocal charge is a weakly clustering linear functional.
\end{corol}

\subsection{Right and two-sided pseudolocal charges}\label{rtquasi}

Let $Q=\{n\mapsto Q_n\}$ be a pseudolocal sequence with respect to $\omega$. We note that the limit $\lim_{n\to\infty} (\omega(A^\star Q_n)-\omega(A^\star)\omega(Q_n)) = \lim_{n\to\infty} \overline{(\omega(Q_n^\star A)-\omega(Q_n^\star)\omega(A))}=\overline{Q_\omega(A)}$ exists. We define the right pseudolocal charge $Q_\omega^\star$ as
\beq\label{Qcc}
	Q_\omega^\star(A^\star) = \overline{Q_\omega(A)}
\eeq
for all $A\in\obs$, and denote the resulting set of linear functionals by $\obsq_\omega^\star$. By definition, there is a bijection $\obsq_\omega\to\obsq_\omega^\star$, induced by the $\star$ involution\footnote{If $Q^\star=\{n\mapsto Q_n^\star\}$ is also a pseudolocal sequence, then there is also the left pseudolocal charge $(Q^\star)_\omega$, not to be confused with the right pseudolocal charge $Q^\star_\omega$.}. Note that the map from pseudolocal sequences to right pseudolocal charges $Q\mapsto Q_\omega^\star$ is linear (not anti-linear).

Let $\omega$ be a clustering state, and recall the form $\bra\cdot,\cdot\ket_\omega$, the subspace $\obsd_\omega$ on which it is degenerate, and the quotient space $\obsn_\omega:=\obs/\obsd_\omega$. We may define another positive semidefinite sesquilinear form on $\obs$ by
\beq\label{defstar}
	\bra A,B\ket_\omega^\star := \bra B^\star,A^\star\ket_\omega,\quad A,B\in\obs.
\eeq
Note that $\bra A,B\ket^\star_\omega = \sum_{x\in\Z^\di} (\omega(B_x A^\star) - \omega(B)\omega(A^\star))$. This is degenerate on the subspace $\obsd^\star_\omega := \{A^\star:A\in\obsd_\omega\}$ (which may be different from $\obsd_\omega$), and Cauchy completing the set of local elements $\obsn^\star_\omega:=\obs/\obsd^\star_\omega$ with respect to the norm induced by this form we obtain a Hilbert space $\obsh^\star_\omega$ on which $\bra\cdot,\cdot\ket^\star$ is an inner product. Clearly, $\obsh_\omega^\star = (\obsh_\omega)^\star$.

It is a simple matter obtain the equivalent result to Theorem \ref{propoleft}, that there is an anti-linear bijection $\dens^\star:\obsq_\omega^\star\to\obsh^\star_\omega$ giving the densities of the right pseudolocal charges.
\begin{theorem}\label{proporight}
Let $\omega$ be a clustering state on $\obs$. There exists a bijection $\dens^\star:\obsq_\omega^\star\to\obsh_\omega^\star$ such that, for every $Q_\omega^\star\in\obsq_\omega^\star$ and $A\in\obs$,
\beq\label{Qomright}
	Q_\omega^\star(A) = \bra \dens^\star(Q_\omega^\star),A\ket_\omega^\star.
\eeq
In particular, $Q_\omega^\star$ can be extended to a continuous linear functional on $\obsh_\omega^\star$.
\end{theorem}
\proof We use the $\star$ bijection $\obsq_\omega^\star\to\obsq_\omega$ and the equality $\obsh_\omega^\star = (\obsh_\omega)^\star$ in order to define the bijection $\dens^\star$ as
\beq\label{defds}
	\dens^\star(Q_\omega^\star) = \dens(Q_\omega)^\star.
\eeq
Using \eqref{Qcc}, \eqref{defstar} and Theorem \ref{propoleft}, we verify that it satisfies \eqref{Qomright}. \eproof

Recall that any convex, real linear combination of positive semidefinite sesquilinear forms also is a such a form. Let us define
\beq\label{dh}
	\dbra A,B\dket_\omega := \frc12\big(\bra A,B\ket_\omega + \bra A,B\ket_\omega^\star\big),\quad A,B\in\obs.
\eeq
Clearly, this is degenerate on $\obsdd_\omega:=\obsd_\omega\cap \obsd^\star_\omega$, and we may complete the quotient space $\obsdn_\omega:=\obs/\obsdd_\omega$ to a Hilbert space $\obsdh_\omega$.
\begin{lemma} \label{lemstar} The following statements hold:
\bi
\item[(i)] $(\obsdh_\omega)^\star = \obsdh_\omega$.
\item[(ii)] If a sequence $j\mapsto A_j\in\obs$ Cauchy converges with respect to $||\cdot||_{\obsdh_\omega}$, then it does so both with respect to $||\cdot||_{\obsh_\omega}$ and $||\cdot||_{\obsh_\omega^\star}$, as does the sequence $j\mapsto A_j^\star$.
\ei
\end{lemma}
\proof Since $\dbra A,A\dket_\omega = \bra A,A\ket_\omega + \bra A^\star,A^\star\ket_\omega$ the first statement follows. From the fact that the linear combination in \eqref{dh} is convex, a sequence in $\obs$ is Cauchy with respect to the norm $||\cdot||_{\obsdh_\omega}$ if and only if it is Cauchy both with respect to the norms $||\cdot||_{\obsh_\omega}$ and $||\cdot||_{\obsh_\omega^\star}$, which implies the second statement. \eproof

There are natural linear maps $\qmap:\obsdn_\omega\to \obsn_\omega$ and $\qmap^\star:\obsdn_\omega\to \obsn^\star_\omega$ defined by inclusion of cosets, with in particular
\beq\label{qmapdh}
	\bra\qmap(A),\qmap(B)\ket_\omega = \bra A,B\ket_\omega,\quad
	\bra\qmap^\star(A),\qmap^\star(B)\ket^\star_\omega = \bra A,B\ket_\omega^\star
\eeq
for all $A,B\in\obsdn_\omega$. These maps are both surjective but not necessarily injective (unless $\obsd_\omega = \obsd^\star_\omega$). Under the $\star$ involution, they satisfy
\beq\label{qmapstar}
	\qmap(A)^\star= \qmap^\star(A^\star)
\eeq
for every $A\in\obsdn_\omega$. By Lemma \ref{lemstar}, these maps extend to continuous linear maps $\qmap:\obsdh_\omega\to \obsh_\omega$ and $\qmap^\star:\obsdh_\omega\to \obsh^\star_\omega$ (which are, in general, neither surjective nor injective). By the Reisz representation theorem, there is a unique continuous linear map $\map:\obsh_\omega\oplus \obsh_\omega^\star \to\obsdh_\omega$ (for simplicity we keep its dependence on $\omega$ implicit) such that\footnote{The norm on $\obsh_\omega\oplus \obsh_\omega^\star$ may be taken, as usual, as $A\oplus B \mapsto \sqrt{\bra A,A\ket_\omega +\bra B,B\ket^\star_\omega}$.}
\beq\label{mapM}
	\frc12\big(\bra A,\qmap(\cdot)\ket_\omega + \bra B,
	\qmap^\star(\cdot)\ket_\omega^\star\big)
	= \dbra \map(A,B),\cdot\dket_\omega
\eeq
for all $A\in\obsdh_\omega$ and $B\in\obsdh_\omega^\star$. Note that by \eqref{dh} and \eqref{qmapdh} and by continuity,
\beq\label{mapplus}
	\map(\qmap(A),\qmap^\star(A)) =A
\eeq
for all $A\in\obsdh_\omega$.

Let $Q=\{n\mapsto Q_n\}$ be a pseudolocal sequence, with the property that $Q^\star=\{n\mapsto Q_n^\star\}$ is also pseudolocal. Then we refer to the sum
\beq\label{defts}
	\widehat Q_\omega:=\frc12\big(Q_\omega\circ \qmap_\omega+(Q^\star)_\omega^\star\circ \qmap_\omega^\star\big)
\eeq
as a two-sided pseudolocal charge, the set of which we denote by $\obsdq_\omega$. Below, without additional denomination, ``pseudolocal charge'' will refer to the two-sided version. By the above, this is a continuous linear functional on $\obsdh_\omega$, and in particular for every $A\in\obs$,
\beq
	\widehat Q_\omega(A) = \frc12\big(\bra \dens(Q_\omega),A\ket_\omega
	+\bra \dens^\star(Q_\omega^\star),A\ket_{\omega}^\star\big)
	= \lim_{n\to\infty}\lt(\frc12\omega(\{Q_n^\star,A\})-
	\omega(Q_n^\star)\omega(A)\rt)
\eeq
where $\{\cdot,\cdot\}$ is the anti-commutator. Note that $Q\mapsto \widehat Q_\omega$ is anti-linear.

The above allows us to show that two-sided pseudolocal charges are in bijection with the Hilbert space $\obsdh_\omega$, seen again as the space of their densities.
\begin{theorem}\label{propotwosided}
Let $\omega$ be a clustering state on $\obs$. There exists a bijection $\widehat\dens:\obsdq_\omega\to\obsdh_\omega$ such that, for every $\widehat Q_\omega\in\obsdq_\omega$ and $A\in\obs$,
\beq\label{Qomtwo}
	\widehat Q_\omega(A) =
	\dbra\, \ddens(\widehat Q_\omega),A\dket_\omega.
\eeq
In particular, $\widehat Q_\omega$ can be extended to a continuous linear functional on $\obsdh_\omega$.
\end{theorem}
\proof By definition, for every pseudolocal sequence $Q$ such that $Q^\star$ is also pseudolocal, the functional $\widehat Q_\omega$ is in $\obsdq_\omega$, and every element of $\obsdq_\omega$ is of that form. For every such $Q$, we define
\beq
	\ddens(\widehat Q_\omega) = \map(\dens(Q_\omega),
	\dens^\star((Q^\star)_\omega^\star)).
\eeq
By \eqref{defts}, \eqref{mapM}, and Theorems \ref{propoleft} and \ref{proporight}, this satisfies \eqref{Qomtwo}. The map $\ddens$ is injective by definition, as if $\widehat Q_\omega\neq \widehat Q'_\omega$, then they are different on at least one element of $\obsdh_\omega$, hence $\ddens(\widehat Q_\omega)\neq\ddens(\widehat Q_\omega')$ by \eqref{Qomtwo}. It is also surjective. Indeed, for any $A\in\obsdh_\omega$, let $A_j\in\obs$ be a sequence that Cauchy converges in the norm in $\obsdh_\omega$ with limit $A$ (this exists by the fact that the set $\obsdn_\omega$ is dense in $\obsdh_\omega$). Then by Lemma \ref{lemstar} $A_j$ also converges in $\obsh_\omega$, with limit $\qmap(A)$. Recall that we may construct a pseudolocal sequence $Q=\{n\mapsto Q_n\}$ where $Q_n$ is defined using the $A_j$'s as in Equations \eqref{prQnj} and \eqref{prQndef} in the proof of Theorem \ref{propoleft}. The left pseudolocal charge $Q_\omega$ has density 
\beq\label{prtse1}
	\dens(Q_\omega)=\qmap(A).
\eeq
By Lemma \ref{lemstar}, we also have $A^\star\in\obsdh_\omega$. This may be obtained as a limit of $A_j^\star$ in $\obsdh_\omega$, and thus also $A_j^\star$ converges in $\obsh_\omega$ with limit $\qmap(A^\star)$. Since \eqref{prQnj} and \eqref{prQndef} are $\star$-linear, then the sequence $n\mapsto Q_n^\star$ is also pseudolocal, and the corresponding left pseudolocal charge $(Q^\star)_\omega$ has density
\beq\label{prtse2}
	\dens((Q^\star)_\omega)=\qmap(A^\star).
\eeq
Therefore $Q_\omega$ has a kernel $Q$ such that $Q^\star$ is also pseudolocal, thus $\widehat Q_\omega$ may be defined as in \eqref{defts} and is a two-sided pseudolocal charge. Further, by \eqref{defds}, \eqref{prtse2} and \eqref{qmapstar}, we have $\dens^\star((Q^\star)_\omega^\star) = \dens((Q^\star)_\omega)^\star = \qmap(A^\star)^\star = \qmap^\star(A)$. From \eqref{prtse1} we then obtain
\beq
	\ddens(\widehat Q_\omega) = \map(\dens(Q_\omega),\dens^\star((Q^\star)_\omega^\star)) = \map(\qmap(A),\qmap^\star(A))=A
\eeq
where in the last equality we used \eqref{mapplus}. Thus, given $A\in\obsdh_\omega$ we have found $\widehat Q_\omega\in\obsdq_\omega$ such that $\ddens(\widehat Q_\omega) = A$, showing surjectivity.
\eproof

\begin{rema} The condition that $Q_\omega$ possess a kernel $Q$ such that $Q^\star$ is also pseudolocal is equivalent to the condition that there exists a sequence $j\mapsto Q_j$ of local sequences reproducing $Q_\omega$ as in Theorem \ref{propolocseqleft}, such that the sequence $j\mapsto Q_j^\star$ of local sequences also reproduce a left pseudolocal charge, which is then equal to $(Q^\star)_\omega$.
\end{rema}

The analogues of Theorem \ref{propolocseqleft} and Corollary \ref{coroleft} are immediate:
\begin{theorem} \label{propolocsecrightts}
Let $j\mapsto A_j\in\obs$ be Cauchy sequence with respect to $||\cdot||_{\obsh_\omega^\star}$ ($||\cdot||_{\obsdh_\omega}$) and $A\in\obsh^\star_\omega$ ($A\in\obsdh_\omega$) the corresponding limit. Then the right (two-sided) pseudolocal charge $Q_\omega^\star = (\dens^\star)^{-1}(A)$ ($\widehat Q_\omega = \ddens^{-1}(A)$) may be obtained as the strongly converging limit of the local charges,
\beq
	Q_\omega^\star = \lim_{j\to\infty}
	\lt(\sum_{x\in\Z^\di} \iso_x(A_j)\rt)_\omega^\star,\quad
	\widehat Q_\omega = \lim_{j\to\infty}
	\widehat{\lt(\sum_{x\in\Z^\di} \iso_x(A_j)\rt)}_\omega.
\eeq
\end{theorem}

\begin{corol} \label{cororightts} Right and two-sided pseudolocal charges are weakly clustering linear functionals.
\end{corol}

\subsection{``Lie'' action of pseudolocal charges}

Finally, to any local sequence $Q=\sum_x \iso_x(A)$ we may associate a linear map $\rep Q:\obs\to\obs$, informally given by the commutator $\rep Q(A) = QA-AQ$, and more precisely defined by
\beq\label{lieact}
	\rep Q (B) =  \sum_{x\in\Z^\di} \big(\iota_x(A)\, B
	- B\, \iota_x(A)\big)
\eeq
for all $B\in\obs$ (note that the sum is finite, hence the result is in $
\obs$). Clearly this is a derivation on $\obs$,
\beq\label{der}
	\rep Q(BC) = \rep Q(B) C + B\rep Q(C)\quad\forall\quad B,C\in\obs.
\eeq

Although the following will not be directly used in establishing the main results in the next section, we believe it is an interesting construction that may find applications later on. Consider the continuous linear map
\beq\label{mapM2}
	\map^-:\obsdh_\omega\to\obsdh_\omega,\quad
	\map^-(A) = \map(\qmap(A),-\qmap^\star(A)).
\eeq
As there is no ambiguity possible, we will also use the notation $\map^-(A)$ for $A\in\obs$. Because of \eqref{mapM} and \eqref{mapM2}, the average in $\omega$ of the adjoint action can be written in terms of the map $\map^-$:
\beq
	\omega(\rep Q(B)) = \dbra \map^-(A),B
	\dket_\omega.
\eeq
The result can thus be extended to $B\in\obsdh_\omega$, and by Theorem \ref{propotwosided} there is a pseudolocal charge, $\widehat\dens^{-1}(\map^-(A))$, that reproduces the action $\dbra \map^-(A),\cdot\dket_\omega$ on $\obsdh_\omega$. Further, by continuity of $\map^-$ the density $A$ can also be extended to $\obsdh_\omega$. Again by Theorem \ref{propotwosided} these densities are in bijective correspondence with two-sided pseudolocal charges, and we have a continuous linear map
\beq\label{repdef}
	\rep_\omega:\obsdq_\omega\to\obsdq_\omega,\quad
	\rep_\omega  := \widehat\dens^{-1}\circ \map^-\circ \widehat\dens
\eeq
with
\beq
	\rep_\omega\widehat Q_\omega(B) = \dbra \map^-(\,\widehat\dens(\widehat Q_\omega)),B\dket_\omega\quad \forall\; B\in\obsdh_\omega.
\eeq
This reproduces, in the local case, the action of the derivation $\rep Q$ evaluated in the state $\omega$:
\beq\label{replocal}
	\rep_\omega \widehat Q_\omega(B) = \omega (\rep Q(B))
\eeq
for all $B\in\obs$ if $Q$ is a local sequence. Note that in general, by Corollary \ref{cororightts}, $\rep_\omega\widehat Q_\omega$ satisfies a clustering property akin to the derivation property of $\rep Q$.

\begin{rema} It is simple to extend the above to the adjoint action $\ad Q(Q')$ of local sequences on local sequences, and thus construct a Lie algebra structure on the set of local sequences. This, however, does not immediately extends to a structure on pseudolocal charges.
\end{rema}

\subsection{Pseudolocal flows}

Consider a one-parameter family $\{\omega\}=\{\omega_s:s\in[0,1]\}$ of clustering states. Suppose that for every $A\in\obs$, the function $s\mapsto \omega_s(A)$ is analytic in a neighbourhood of some point $s$. If the clustering property holds uniformly on this neighbourhood of $s$, then it is simple to show that the derivative satisfies
\beq
	\lim_{\dist(A,B)\to\infty}
	\frc{\dd}{\dd s}\omega_{s}(AB) =
	\omega_{s}(B)\frc{\dd}{\dd s}\omega_{s}(A)
	+\omega_{s}(A)\frc{\dd}{\dd s}\omega_{s}(B).
\eeq
Thanks to Corollaries \ref{coroleft} and \ref{cororightts}, pseudolocal charges also satisfy this property. Hence, one may expect that certain subspaces of pseudolocal charges generate vector fields in neighbourhoods of clustering states, and give rise to a Riemannian manifold structure on spaces of such states, with metric tensor built out of the inner product $\bra\cdot,\cdot\ket_\omega$ or its right or two-sided relatives. We will not develop here the details of this manifold structure as this is beyond the scope of this paper. However, we may still define the concept of flows along a path determined by pseudolocal charges, inspired by the manifold analogy.

\begin{defi}\label{defiflow}
Let $\{\omega\}=\{\omega_s:s\in[0,1]\}$ be a one-parameter family of clustering states. If there exists a one-parameter family $\{Q\}=\{Q_{s}\in\obsq_{\omega_s}:s\in[0,1]\}$ of uniformly bounded left pseudolocal charges such that, for every $A\in\obs$, the function $s\mapsto Q_s(A)$ is Lebesgue integrable on $[0,1]$ (with respect to the Borel $\sigma$-algebra) and
\beq\label{flow}
	\omega_{s_+}(A)-\omega_{s_-}(A) = 
	\int_{s_-}^{s_+} \dd s\, Q_{s}(A)\quad\forall \quad 0\leq s_-<s_+\leq 1,
\eeq
then we say that $\{\omega\}$ is a left pseudolocal flow from $\omega_0$ to $\omega_1$ along $\{Q\}$.  If, on $s\in[0,1]$, the states $\omega_s$ are uniformly (sizably) clustering, then we say that the pseudolocal flow is uniformly (sizably) clustering. We define similarly right and two-sided pseudolocal flows.
\end{defi}
Recall that uniform boundedness of $Q_s$ is equivalent to uniform boundedness of the densities $\dens(Q_s)$ by \eqref{normQ} (this holds in the left, right and two-sided versions). Without additional denomination, ``pseudolocal flow'' will refer to the two-sided version.

The above is a somewhat weak definition. One may require additional properties, such as differentiability or analyticity in $s$, whereby the relation $\dd \omega_s(A)/\dd s = Q_s(A)$ makes $\{\omega\}$ into a path whose tangent is determined by $\{Q\}$. In any case, many questions may be brought up: for instance, given a family of pseudolocal sequences and a state $\omega$, is there a pseudolocal flow starting from $\omega$ and going along the associated pseudolocal charges? If it exists, is it unique? What general conditions guarantee existence and uniqueness? We will not address these questions, and rather concentrate on assessing if there exist pseudolocal flows between states of interest.

\section{Time evolution and generalized thermalization} \label{secttime}

In this section we specify $\obs$ to the inductive limit $\bigotimes_{n\in\Z^\di}\obs_{\{n\}}$ of isomorphic finite-dimensional spaces of endomorphisms with the operator norm, or any unital subalgebra thereof that is invariant under Lie actions \eqref{lieact}. This satisfies all properties of the general setup of Section \ref{sectalg} and, as will be clear below, we may define local homogeneous dynamics on it. We note that on $\obs$ there is a natural trace state $\Tr_\obs:=\bigotimes_{n\in\Z^\di} \Tr_{\obs_{\{n\}}}$. It can be extended to $\obsc$ by continuity, and we will denote this extension by $\Tr_\obsc$ (which we normalize such that $\Tr_\obsc(\1)=1$). This has the physical interpretation of an infinite-temperature state.

Below, the phrase ``local Hamiltonian'' will refer to a local sequence $H=\sum_x  \iso_x(h)$ with density $h\in\obs$ satisfying $h^\star=h$.

Using the notion of pseudolocal flows, we propose a natural family of states: that of pseudolocal states. For convenience, we also propose variants of it: weak and analytic pseudolocal states. We will show that the family of pseudolocal states is preserved by finite-time evolution (but, as will be clear from the proof, that of weak pseudolocal states is not). Within this framework, we then propose a definition of (weak, analytic) generalized Gibbs ensembles, which intuitively agrees with the notion used extensively in the literature \cite{rigol2}. Using the results of Kliesch et.~al. \cite{KGKRE}, we will show that any thermal Gibbs state associated to a local Hamiltonian is an analytic GGE, whenever the temperature is higher than a certain universal temperature. Further, we will show that a weak analytic GGE with respect to a local Hamiltonian that does not possess conserved pseudolocal charges other than itself, is a thermal Gibbs state with respect to this Hamiltonian. These two facts provide compelling justifications for our definition of GGEs, and the latter fact is somewhat in the spirit of the eigenstate thermalization hypothesis. We will then show that under certain conditions on the long-time limit, pseudolocal states evolve towards GGEs, thus rigorously showing, within this framework, the emergence of GGEs in non-equilibrium time evolution.

The logical path is as follows: Theorem \ref{theothermal} shows that thermal Gibbs states (at high enough temperatures) are GGEs, and Theorem \ref{theoGGEgibbs} shows how the absence of local conserved charges other than $H$ guarantees that a GGE with respect to $H$ is a thermal Gibbs state. These show that our definition of GGEs connects well with the standard notion of Gibbs states, at least at high enough temperatures. Theorem \ref{propotime} then shows that the structures we have developed, in particular the Hilbert space $\obsdh_\omega$, interact will with time evolution. This is nontrivial, as time evolution defined on $\obsc$ does not immediately implies a well defined time evolution on $\obsdh_\omega$. Theorem \ref{theowb} uses this in order to show that the family of pseudolocal states is stable under time evolution. This further emphasizes the naturalness of pseudolocal states -- this is important, as we have not connected, in general, pseudolocal states with any of the known and well studied states, for instance the equilibrium states based on (generically nonlocal) interactions, although such a connection might exist. Theorem \ref{theoGGE} finally provides the main result. It again uses Theorem \ref{propotime}, for the proof and for making the assumptions weaker (by making the requirement of uniformity of clustering nontrivial just in the infinite-time limit), and, in a sense, it completes Theorem \ref{theowb} to the infinite-time limit.

\subsection{Pseudolocal states and generalized Gibbs ensembles}

\begin{defi}\label{defistates} {\em (Pseudolocal states)}
A pseudolocal state is a state $\omega$ such that there exists $p>\di$ and a uniformly sizably $p$-clustering pseudolocal flow $\{\omega_s:s\in[0,1]\}$ from the infinite-temperature state, $\omega_0=\Tr_{\obsc}$, to that state, $\omega_1=\omega$.

A pseudolocal state is weak if the flow is uniformly clustering but not necessarily sizably $p$-clustering.

A pseudolocal state (weak pseudolocal state) is analytic if there exists a neighbourhood $K$ of $[0,1)$ such that for every $A\in\obs$ the function $s\mapsto\omega_s(A)$ has an analytic extension to $K$, and $\{\omega_s:s\in K\}$ is a uniformly sizably $p$-clustering (uniformly clustering) family of linear functionals for every compact subset of $K$.
\end{defi}
Here and below, a neighbourhood is a simply connected open set.

Above we made use, implicitly, of two-sided pseudolocal flows in order to define pseudolocal states. This is natural as conjugation properties of the state translate into $\star$-invariant properties of the pseudolocal charges: the fact that $\overline{\omega_s(A)} = \omega_s(A^\star)$ for all $A\in\obs$ and almost all $s$ implies that, if $Q_s$ is a kernel for $\widehat Q_s$, then $\widehat{(Q^\star_s)}_{\omega_s} = \widehat{Q}_s$ (for almost all $s$); equivalently, if $B_s\in\obsdh_{\omega_s}$ is the density of $\widehat Q_s$, then $B_s^\star = B_s$  (for almost all $s$).

\begin{defi} \label{defiGGE} {\em (Generalized Gibbs ensembles, GGEs)}
A (weak, analytic) GGE with respect to a local hamiltonian $H$ is a (weak, analytic) pseudolocal state $\omega$, with  flow $\{\omega_s:s\in[0,1]\}$ and pseudolocal charges $\{\widehat Q_s:s\in[0,1]\}$, with the property that for almost all $s\in[0,1]$, we have $\omega_s(\rep H(\obs)) = \{0\}$ and $\widehat Q_s(\rep H(\obs))=\{0\}$.
\end{defi}
As we recall below, the Lie action $\ii\rep H(A)$ represents the infinitesimal time evolution of $A$ with respect to $H$. Therefore, a GGE with respect to $H$ is interpreted as a pseudolocal state whose flow is conserved by $H$, and determined by conserved pseudolocal charges. Note that, of course, if the GGE is analytic, the condition that for almost all $s\in[0,1]$, we have $\omega_s(\rep H(\obs)) = \{0\}$, implies the same equation for all $s$ in the analyticity region, and implies $\widehat Q_s(\rep H(\obs))=\{0\}$ for all $s\in[0,1]$.

\begin{defi}\label{defimixing} {\em (Completely mixing)} A local hamiltonian $H$ is completely mixing if for every clustering state $\omega$ such that $\omega(\rep H(\obs))=\{0\}$, it holds $\{\widehat Q\in\obsdq_\omega:\widehat Q(\rep H(\obs))=\{0\}\} = \C\widehat H_\omega$.
\end{defi}
In other words, a completely mixing local Hamiltonian $H$ does not possess conserved pseudolocal charges other than those having kernels that are scalar multiples of $H$ itself.

Let $H = \sum_x \iso_x(h)$ (with $h\in\obs$ satisfying $h^\star=h$). Consider high enough temperatures $\beta^{-1}$. For every $A\in\obs$, the following limit exists
\beq\label{limthermal}
	\omega_\beta^{\rm th}(A) = \lim_{n\to\infty} \omega^{{\rm th},n}_\beta(A),\quad  \omega^{{\rm th},n}_\beta(A) := \frc{
	\Tr_{\obs_{\ball(n)}}\lt(e^{-\beta \sum_{x\in\ball(n)} \iso_x(h)}\,A\rt)}{
	\Tr_{\obs_{\ball(n)}}\lt(e^{-\beta \sum_{x\in\ball(n)} \iso_x(h)}\rt)}
	\quad (\supp(A)\in\ball(n)),
\eeq
and its extension to $\obsc$ by continuity defines a thermal Gibbs state at inverse temperature $\beta$ with respect to $H$; the state is translation invariant, exponentially clustering, and the function $\omega_\beta^{\rm th}(A)$ is analytic in $\beta$ for every $A\in\obs$ (see \cite{Ruelle,BR1,Simon} and references therein). Let $\beta_0\in\R$. Let $\gamma = [0,\beta_0]$ if $\beta_0>0$ and $\gamma=[\beta_0,0]$ if $\beta_0<0$. If the analytic continuation of $\omega_{\beta}^{\rm th}(A)$ on a simply connected region $K\supset \gamma$ exists for every $A\in\obs$ and defines, for $\beta\in\gamma$, translation-invariant states $\omega_{\beta}^{\rm th}$ on $\obsc$; and if $p$-clustering holds for every $p>\di$ uniformly on compact subsets of $K$; then we will refer to $\omega_{\beta_0}^{\rm th}$ as a {\em high-temperature  Gibbs state} at temperature $\beta_0$. Note that clustering implies the absence of phase transitions on the whole path $\gamma$.

The goal of the next theorem is to establish that high-temperature  Gibbs states, as defined above, are (weak) analytic GGEs with respect to $H$ as per Definition \ref{defiGGE}. In order to be precise, we use the specific results of \cite{KGKRE} (and the fundamental result of \cite{Araki69} in one dimension), which, we have found, are especially convenient for our purpose. In particular, we consider the following inverse temperature bound \cite{KGKRE} guaranteeing appropriate clustering and analyticity:
\beq
	\beta_*:=\lt\{\ba{ll}
	\frc1{2||h||}\log\big[\frc{1+\sqrt{1+2/(\di e)}}2\big]& (\di>1) \\
	\infty & \di=1.
	\ea\rt.
\eeq
We note that the limit \eqref{limthermal} exists and defines a translation-invariant linear functional on $\obs$ for any $\beta\in \C$, $|\beta|<\beta_*$ by the results \cite[Thm 2.1]{Araki69} for $\di=1$, and \cite[Corol 2]{KGKRE}\footnote{In particular, this corollary implies that $\omega^{{\rm th},n}_\beta(A)$ form, for $n\in\N$, a Cauchy sequence.} for $\di>1$; for $\beta\in\R$, $|\beta|<\beta_*$, this therefore defines a state. The following theorem establishes that $\omega_\beta^{\rm th}$ is an analytic GGE with respect to $H$ for every $|\beta|<\beta_*$. It is then a simple matter to see that these are high-temperature Gibbs state, and that in general, high-temperature Gibbs states are weak analytic GGEs (in general no condition on sizability of clustering), and points (ii) and (iv) of the following theorem also hold generally for such states (with $K$ the appropriate analyticity region).
\begin{theorem} \label{theothermal}
Let $K:=\{\beta\in\C,\,|\beta|<\beta_*\}$, and let $J$ be a compact subset of $K$. 
\bi
\item[(i)] The family of linear functionals $\{\omega_\beta^{\rm th}:\beta\in J\}$ is uniformly sizably $p$-clustering for any $p>\di$.
\item[(ii)] For every $A\in\obs$, the function $\omega_\beta^{\rm th}(A)$ of $\beta$ is analytic on $K$, and can be expressed as
\beq
	\omega_\beta^{\rm th}(A) = \Tr_{\obsc}(A)-
	\int_0^\beta\dd \alpha\,\dbra h,A\dket_{\omega_{\alpha}^{\rm th}}
\eeq
where the integral lies in $K$.
\item[(iii)] For every $\beta\in \R$, $|\beta|<\beta_\star$, the state $\omega_\beta^{\rm th}$ is an analytic GGE with respect to $H$.
\item[(iv)] There exists $\nu>0,\,a>0$ such that, for every $A\in\obs$ with $\siz{A}<\ell$ and every $\beta\in J$,
\beq
	|\omega^{\rm th}_\beta(A)| < (1+\nu\ell^a)\,||A||.
\eeq
\ei
\end{theorem}
\proof In the case $\di=1$, results of Araki \cite{Araki69} and others (see \cite[Thm 2.1, 2.3]{Araki69}, \cite[Thm III.2]{GN98} and \cite[Thm 1.3]{Mat02}) may be used in order to establish some aspects of this theorem. We instead directly use results of Kliesch {\em et. al.} \cite{KGKRE} valid for all $\di\geq 1$. There, it is shown that \cite[Thm 2]{KGKRE} (here we present a version sufficient for our purposes) for every $n\in \N$, $\beta\in K$, $\ell>0$, $A,B\in\obs$ with $\siz{A},\siz{B}<\ell$, and $\dist(A,B)>L_0$,
\beq\label{proti}
	|\omega_\beta^{{\rm th},n}(AB)-\omega_\beta^{{\rm th},n}(A)\omega_\beta^{{\rm th},n}(B)| \leq \nu \ell ||A||\,||B||\,e^{-\dist(A,B)/\xi}.
\eeq
The exponent $\xi>0$ as well as the bounding constant $\nu$ only depend on the norm of the Hamiltonian density $||h||$, the lattice dimensionality $\di$ and the temperature $\beta$, and can be chosen uniform on any compact subset of $|\beta|<\beta_*$ and for all $n>0$. The minimal length $L_0$ depends on the same quantities (uniformly on compact subsets of $|\beta|<\beta_*$ and for all $n>0$), as well as on $\ell$, and can be chosen of the form $L_0 = a\log(\ell) + L_0'$ where $a>0$ and $L_0'>0$ do not depend on $\ell$. In order to make contact with Definition \ref{defisiz} of sizable clustering, we need to take away the condition $\dist(A,B)>L_0$. In the cases $\dist(A,B)\leq L_0$, we may simply bound the left-hand side of \eqref{proti} by $2||A||\,||B||$. Otherwise, we multiply the right-hand side of \eqref{proti} by $e^{L_0/\xi}$. This amounts, up to an $\ell$-independent factor that is uniform on compact subsets of $|\beta|<\beta_*$, to multiplying by $\ell^{a/\xi}$, and thus, now without the condition on $\dist(A,B)$, we have  (with an appropriate re-definition of $\nu$)
\beq\label{clusprot}
	|\omega_\beta^{{\rm th},n}(AB)-\omega_\beta^{{\rm th},n}(A)\omega_\beta^{{\rm th},n}(B)| \leq \nu \ell^{a/\xi+1} ||A||\,||B||\,e^{-\dist(A,B)/\xi}.
\eeq
We may take the the infinite-$n$ limit, and therefore the linear functional $\omega_\beta^{\rm th}$ is uniformly sizably exponentially clustering for $\beta$ on every compact subset of $K$. This shows point (i).


Further, it is clear that, for every $n\in\N$, $\beta\in\C$ and $A\in\obs$,
\beq\label{omevbtg}
	\frc{\dd}{\dd\beta} \omega_\beta^{{\rm th},n}(A)
	= -\sum_{x\in\ball(n)}\lt( \omega_\beta^{{\rm th},n}(\iso_x(h)A)
	- \omega_\beta^{{\rm th},n}(\iso_x(h))
	\omega_\beta^{{\rm th},n}(A)\rt).
\eeq
A similar expression holds with the order of $\iso_x(h)$ and $A$ inverted. Symmetrizing and integrating,
\beq
	\omega_\beta^{{\rm th},n}(A)
	= \Tr_\obsc(A)-\int_0^\beta \dd \alpha\,
	\sum_{x\in\ball(n)}\lt( 
	\frc12 \omega_\alpha^{{\rm th},n}(\{\iso_x(h),A\})
	- \omega_\alpha^{{\rm th},n}(\iso_x(h))
	\omega_\alpha^{{\rm th},n}(A)\rt).
\eeq
With $n$ large enough, we may find $m$ large enough such that $\supp(\iso_x(h))\cap\supp(A)=\emptyset$ for all $x\in \ball(n)\setminus\ball(m)$. With such a choice, \eqref{clusprot} implies that for every $A\in\obs$ and $\alpha\in K$ there exists $\t\nu>0$, uniform for $\alpha$ in compact subsets of $K$, such that
\beq
	\sum_{x\in\ball(n)\setminus\ball(m)}\Big|
	\frc12 \omega_\alpha^{{\rm th},n}(\{\iso_x(h),A\})
	- \omega_\alpha^{{\rm th},n}(\iso_x(h))
	\omega_\alpha^{{\rm th},n}(A)\Big|
	\leq \t\nu \sum_{d=m}^\infty \are_\di(d) e^{-d/\xi}.
\eeq
Therefore,
\beq\label{omevbth}
	\omega_\beta^{{\rm th},n}(A)
	= \Tr_\obsc(A)-\int_0^\beta \dd \alpha\,
	\sum_{x\in\ball(m)}\lt( 
	\frc12 \omega_\alpha^{{\rm th},n}(\{\iso_x(h),A\})
	- \omega_\alpha^{{\rm th},n}(\iso_x(h))
	\omega_\alpha^{{\rm th},n}(A)\rt) + \delta(m)
\eeq
where $|\delta(m)| \leq  \int_0^\beta \dd\alpha\,\t\nu \sum_{d=m}^\infty \are_\di(d) e^{-d/\xi}$. We may now take the infinite-$n$ limit. By \eqref{clusprot}, the integrand on the right-hand side of \eqref{omevbth} is uniformly bounded for $\alpha$ in compact subsets of $K$ and $n\in\N$, and by the results recalled, the limit $n\to\infty$ exists pointwise in $\alpha$. By Lebesgue's dominated convergence theorem the limit of the integral therefore exists and is the integral of the limit. The result is
\beq\label{omevbth2}
	\omega_\beta^{{\rm th}}(A)
	= \Tr_\obsc(A)-\int_0^\beta \dd \alpha\,
	\sum_{x\in\ball(m)}\lt( 
	\frc12 \omega_\alpha^{{\rm th}}(\{\iso_x(h),A\})
	- \omega_\alpha^{{\rm th}}(\iso_x(h))
	\omega_\alpha^{{\rm th}}(A)\rt) + \delta(m).
\eeq
Clearly, for every $A\in\obs$ and $n\in\N$ the function $\omega^{{\rm th},n}_\beta(A)$ of $\beta$ is analytic in $\C$. Therefore, the integrand on the right-hand side of \eqref{omevbth} is analytic in $\alpha$. Since it is uniformly bounded for $\alpha$ in compact subsets of $K$ and $n\in\N$, by a theorem Stieltjes the limit $n\to\infty$ is analytic in $K$. Therefore the integrand on the right-hand side of \eqref{omevbth2} is analytic for $\alpha\in K$.

Since $\omega^{\rm th}_\alpha$ is uniformly clustering linear functional for $\alpha$ in compact subsets of $K$, Theorem \ref{propoconv} implies that the infinite-$m$ limit of the integrand on the right-hand side of \eqref{omevbth2} exists uniformly in $\alpha$. The result, $\dbra h,A\dket_{\omega^{\rm th}_\alpha}$, is therefore analytic in $\alpha\in K$. Using $\lim_{m\to\infty}\delta(m)=0$, we may then re-write the limit $m\to\infty$ of the right-hand side of \eqref{omevbth} as
\beq\label{prtherm0}
	\omega_{\beta}^{\rm th}(A) = \Tr_\obsc(A)-
	\int_0^\beta\dd \alpha\,
	\dbra h,A\dket_{\omega_{ \alpha}^{\rm th}} \quad (\beta \in K).
\eeq
Since the integrand is analytic in $\alpha$, we have shown that $\omega_\beta^{\rm th}(A)$ is analytic in $\beta$, and point (ii) follows. 

In order to explicitly write the pseudolocal flow, we consider the family of thermal states $\{\omega_s:=\omega_{\beta s}^{\rm th}:s\in [0,1]\}$. Clearly, for $\beta\in\R$,  $|\beta|<\beta_*$, \eqref{prtherm0} may be expressed in terms of the action of the (two-sided) local charge associated to the local sequence $H$,
\beq\label{prtherm1}
	\omega_{u}(A) = \Tr_\obsc(A)-\beta
	\int_0^u\dd s\,
	\widehat H_{\omega_{ s}}(A).
\eeq
By the uniform clustering property of thermal states and Theorem \ref{propoconv}, the quantity $\dbra h,h\dket_{\omega_{ s}}$ is uniformly bounded over $s\in[0,1]$. Thus, $\{\omega_{ s}:s\in[0,1]\}$ is a uniform sizably $p$-clustering pseudolocal flow from the infinite-temperature state to the thermal state $\omega_\beta^{\rm th}$ (for any $p>\di$), showing the first paragraph of Definition \ref{defistates}. The third paragraph of Definition \ref{defistates} follows from points (i) and (ii), and therefore point (iii) is shown.

Finally, using \eqref{prtherm0}, Theorem \ref{propoconv} and the established sizable clustering property (point (i)), there exists $a>0$ such that, for all $A\in\obs$ with $\siz{A}<\ell$,
\beq
	|\omega_\beta^{\rm th}(A)| \leq ||A|| + \int_0^{\beta}
	|\dd \alpha|\, |\dbra h,A\dket_{\omega_\alpha^{\rm th}}|
	\leq (1+|\beta|F \ell^a \siz{A}\,\siz{h}\,||h||)\,||A||,
\eeq
and point (iv) follows.
\eproof

\begin{theorem} \label{theoGGEgibbs} Let $H$ be a completely mixing local Hamiltonian, and let $\omega$ be a weak analytic GGE with respect to $H$, with $\{\omega_s:s\in[0,1]\}$ the associated flow. 
\bi
\item[(i)] Assume that for every $s\in[0,1]$, there exists $A_s\in\obs$, $A_s^\star = A_s$, such that $\widehat H_{\omega_s}(A_s)\neq0$ and $\dd \omega_{s'}(A_s)/\dd s'\Big|_{s'=s}\neq0$. Then $\omega$ is a high-temperature Gibbs state $\omega_{\beta}^{\rm th}$ with respect to $H$ at inverse temperature
\beq
	\beta = -\int_0^{1} \dd s\, \frc{\dd \omega_{s'}(A_s)/\dd s'|_{s'=s}}{
	\widehat H_{\omega_s}(A_s)} \in \R.
\eeq
\item[(ii)] Assume that there exists $A\in\obs$ such that $\widehat H_{\Tr_\obsc}(A)\neq0$ and $\dd \omega_s(A)/\dd s\big|_{s=0}\neq0$. Then there exists a smooth open curve $u\in(0,1)\mapsto \beta_u\in\C$ such that for every $A\in\obs$, the quantity $\omega(A)$ is the limit $u\to1$ along this curve of analytically continued high-temperature Gibbs states $\omega_{\beta}^{\rm th}$ with respect to $H$; that is $\omega(A) = \lim_{u\to1}\omega^{\rm th}_{\beta_u}(A)$. If, additionally, the analyticity region $K$ of the analytic GGE contains the full interval $[0,1]$ and there exists $B\in\obs$ such that $\widehat H_{\omega}(B)\neq0$, then $\lim_{u\to 1}\beta_u$ converges in $\C$.
\ei
\end{theorem}
\proof If $\omega=\Tr_\obsc$ then the Theorem is immediate, hence let us assume otherwise.

We denote by $\{\omega_s:s\in[0,1]\}$ the flow from $\omega_0=\Tr_\obsc$ to $\omega_1=\omega$, by $\{\widehat Q_s:s\in[0,1]\}$ the associated pseudolocal charges, and by $K$ the neighbourhood of $[0,1)$ where $s\mapsto \omega_s(A)$ has an analytic extension and where the family of linear functionals $\{\omega_s:s\in K\}$ is uniformly bounded and clustering on compact subsets.

By the completely mixing property, the family of pseudolocal charges $\{\widehat Q_s:s\in[0,1]\}$ must lie in the one-dimensional space generated by the kernel $H$, so that there exists $s\mapsto\alpha(s)\in\C$ such that $\widehat Q_s = \alpha(s) \widehat H_{\omega_s}$ for all $s\in[0,1]$.

Let $A\in\obs$ and consider
\beq\label{prggetr7}
	 \widehat H_{\omega_s}(A)
	=\lim_{X\to\Z^\di}\sum_{x\in X}\lt(\frc12 \omega_s(\{\iso_x(h),A\})
	-\omega_s(h)\omega_s(A)\rt)
\eeq
(for definiteness, here and below $X\to\Z^\di$ means $X=\ball(n),\,n\to\infty$). Fix an $X$ on the right-hand side of \eqref{prggetr7}. Then the resulting function of $s$ is analytic in $K$. Further, by Theorem \ref{propoconv} and clustering, the limit $X\to\Z^\di$ exists pointwise for $s\in K$, and by using uniform clustering, the sequence of functions of $s$ indexed by $X$ is uniformly bounded on every compact subsets of $K$. Hence by a theorem of Stieltjes, the limit $X\to\Z^\di$ is analytic as a function of $s\in K$.

By analyticity, for every $A\in\obs$ and on $s\in [0,1)$,
\beq\label{prggetr2}
	\frc{\dd}{\dd s} \omega_s(A) = \alpha(s)\widehat H_{\omega_s}(A).
\eeq
By \eqref{prggetr2} and the result established above it, $\alpha(s)$ has an extension that is meromorphic on $K$. 

Choose $A_s\in\obs$ to be those assumed to exist in point (i). Then
\[
	\alpha(s)=\frc{\dd \omega_{s'}(A_s)/\dd s'|_{s'=s}}{\widehat H_{\omega_s}(A_s)}
\]
(which real for real $s$) is finite and nonzero for all $s\in[0,1]$, and therefore $\alpha(s)$ is analytic and nonzero on a neighborhood of $s\in[0,1)$ and left-continuous at 1. We may define the monotonically increasing function
\beq\label{us}
	u(s) := \frc{\int_0^s \dd s'\,\alpha(s')}{\int_0^{s_1} \dd s'\,\alpha(s')}
\eeq
with $s_1=1$ and for $s\in[0,1]$. Defining $\omega_{u(s)}' := \omega_{s}$, we have, for every $A\in\obs$,
\beq\label{prggeu}
	\frc{\dd}{\dd u} \omega_u'(A) = -\beta \widehat H_{\omega_u'}(A),\quad \beta = -\int_0^{s_1} \dd s'\,\alpha(s'),
\eeq
and $\omega_0'=\Tr_\obsc$, $\omega_1'=\omega_{1}$.

On the other hand, choose $A$ to be that assumed to exist in point (ii). Both $\widehat H_{\omega_s}(A)$ and $\dd \omega_s(A)/\dd s$ are analytic functions of $s$ in $K$. Hence, if any has an infinite number of zeros, the zeroes must accumulate to a boundary point of $K$. Therefore there exists $s_1\in(0,1)$ such that there is a neighbourhood $K'\supset [0,s_1]$, $K'\subset K$ on which both functions only have a finite number of zeroes, and we may choose $s_1$ as close as we wish to 1. Hence, we may find a smooth path $\gamma\subset K$ from $0$ to $s_1$, on which $\widehat H_{\omega_s}(A)$ and $\frc{\dd}{\dd s} \omega_s(A)$ are both uniformly nonzero. On a simply connected neighbourhood $L$ of this path, $\alpha(s)$ is analytic and uniformly nonzero. Hence $\alpha(s) = \p g(s)$ where $g(s)$ is conformal on $L$. Define $u(s) \propto g(s)$ by \eqref{us}, where now the integration paths lie in $L$ (and we may choose $s_1$ such that the denominator is nonzero). This satisfies $u(s_1)=1$ and $u(0)=0$, and $u(L)$ is a simply connected Riemann surface (which might be implemented on a multiple cover of $\C$). Then, for every $A\in\obs$, \eqref{prggeu} holds on the Riemann surface $u(L)$, and $\omega_0'=\Tr_\obsc$, $\omega_1'=\omega_{s_1}$.

We show below that \eqref{prggeu} with the above initial condition $\omega_0'$ has a unique solution (a unique family of states $\{\omega_u:u\in u(L)\}$). The solution may be explicitly written as the analytic continuation of a high-temperature Gibbs state $\omega_u = \omega^{\rm th}_{\beta u}$ for every $|u|<\beta_*/|\beta|$, by analyticity (Theorem \ref{theothermal}) and by \eqref{prtherm1}. Thus,  $\omega_1 = \lim_{s_1\to1}\omega_{s_1}$ is either a high-temperature Gibbs state (in the case of point (i)) or an analytic extension thereof (in the case of point (ii)).

If $K$ includes $[0,1]$ and there exists $B\in\obs$ such that $\widehat H_{\omega}(B)\neq0$, then it is clear that $\alpha(s)$ is not singular at $s=1$, and so $\lim_{s_1\to 1}\beta$ exists and is finite.

Uniqueness is proved by recursively constructing the higher-order derivatives. By scaling \eqref{prggeu}, it is sufficient to prove this for \eqref{prggetr2} with $\alpha(s)=1$. For any $B\in\obs$, define the derivation $D_s(B)$ on polynomials of $\omega_s$ by $D_s(B) \omega_s(A) = \frc12 \omega_s(\{B,A\}) - \omega_s(B)\omega_s(A)$. Assume that for all $j\leq k\in\N$, the relation
\beq\label{prggetr5}
	\frc{\dd^j}{\dd s^j}\omega_s(A)
	= \lim_{X_j\to\Z^\di}\cdots \lim_{X_1\to\Z^\di}\sum_{x_1\in X_1}\cdots \sum_{x_j\in X_j}
	D_s(\iso_{x_1}(h))\cdots D_s(\iso_{x_j}(h))\,\omega_s(A)
\eeq
holds at $s=0$ for all $A\in\obs$. Since $D_s(B)$ is a derivation by definition, this implies that the relation holds on any polynomials of $\omega_s$ (at $s=0$). Then we have
\beqa
	\lt.\frc{\dd^{k+1}}{\dd s^{k+1}}\omega_s(A)\rt|_{s=0} &=&
	\frc{k!}{2\pi \ii} \oint_{0} \frc{\dd s}{s^{k+1}} \frc{\dd}{\dd s}
	\omega_s(A)\n
	&=& 
	\frc{k!}{2\pi \ii}\oint_{0} \frc{\dd s}{s^{k+1}}
	\lim_{X\to\Z} \sum_{x\in X}\lt(
	\frc12\omega_s(\{\iso_x(h),A\})
	- \omega_s(h)\omega_s(A)
	\rt)\n
	&=& 
	\lim_{X\to\Z^\di} \sum_{x\in X}
	\frc{k!}{2\pi \ii} \oint_{0} \frc{\dd s}{s^{k+1}}
	\lt(
	\frc12\omega_s(\{\iso_x(h),A\})
	- \omega_s(h)\omega_s(A)
	\rt)\no
\eeqa
where in the third step we used uniform clustering in $K$ (thus in a neighbourhood of 0). Continuing,
\beqa
	&=& 
	\lim_{X\to\Z^\di} \sum_{x\in X}
	\lt.\frc{\dd^k}{\dd s^{k}}
	\lt(
	\frc12\omega_s(\{\iso_x(h),A\}_s)
	- \omega_s(h)\omega_s(A)
	\rt)\rt|_{s={0}}\n
	&=& 
	\lim_{X_{k+1}\to\Z^\di}\cdots \lim_{X_1\to\Z^\di}\sum_{x_1\in X_1}\cdots \sum_{x_{k+1}\in X_{k+1}}
	D_s(\iso_{x_1}(h))\cdots D_s(\iso_{x_k}(h))
	\cdot \n && \cdot
	\lt(
	\frc12\omega_s(\{\iso_{x_{k+1}}(h),A\})
	- \omega_s(h)\omega_s(A)
	\rt)\Big|_{s={0}}\n
	&=& 
	\lim_{X_{k+1}\to\Z^\di}\cdots \lim_{X_1\to\Z^\di}\sum_{x_1\in X_1}\cdots \sum_{x_{k+1}\in X_{k+1}}
	D_s(\iso_{x_1}(h))\cdots D_s(\iso_{x_{k+1}}(h))\, \omega_s(A).\n
	\no
\eeqa
Hence \eqref{prggetr5} holds for $j=k+1$ as well. Therefore any two families $\{\omega_s\}$ and $\{\omega_s'\}$ satisfying the same equation \eqref{prggetr2} (with $\alpha(s)=1$) and the same initial condition are such that $\omega_s(A)$ and $\omega_s'(A)$ have equal derivatives at all orders at $s=0$ for any $A\in\obs$. Since both are analytic, by assumption, on a neighbourhood of $[0,1)$, they are equal on this neighbourhood. Since equality holds for any $A\in\obs$, by continuation it holds for any $A\in\obsc$.
\eproof

\begin{rema} In Theorem \ref{theoGGEgibbs}, point (i) gives a high-temperature Gibbs state at temperature $\beta^{-1}$, so that no phase transition occurs at higher temperature, while point (ii) indicate that a phase transition may occur.
\end{rema}

\begin{rema}\label{remaGGE}
The connection with the usual (un-rigorous) form
\[
	\omega(A) = \frc{\Tr\lt(e^{-\sum_i \beta_i Q_i}\,A\rt)}{
	\Tr\lt(e^{-\sum_i \beta_i Q_i}\rt)}
\]
of GGEs used in the literature, where in the present framework $Q_i$ would be pseudolocal sequences, may be obtained by formal manipulations as follows. We choose a differentiable path $\beta_i(s)$ in the space of potentials, with $\beta_i(1)=\beta_i$ and that reaches the infinite-temperature state $\beta_i(0)=0$ at $s=0$, and then we differentiate with respect to the parameter of the path. The result, $\sum_i \dot\beta_i(s) (\omega(Q_iA) - \omega(Q_i)\omega(A))$ (where $\dot\beta_i(s) = \dd \beta_i(s)/\dd s$), is then interpreted as the action on $A$ of the pseudolocal charge $\widehat Q_s=\sum_i\dot\beta_i(s) \widehat {(Q_i)}_\omega$ (assuming that the series converge). Thus the path is a pseudolocal flow from the infinite-temperature state, and, with appropriate additional conditions (including clustering), the state $\omega$ is a GGE according to the definition proposed above. Note that the expression of $\widehat Q_s$ here may be seen as its decomposition in some basis (from the Hilbert space structure).
\end{rema}

\subsection{Time evolution}

Let the local Hamiltonian $H$ be a local sequence with density $h\in\obs$ such that $h^\star = h$. For every $A\in\obs$, consider the power series in $t$ (the evolution of $A$ by the Hamiltonian $H$ for a time $t$) given by
\beq\label{time}
	\tau_t(A) := e^{\ii t\rep H}(A) =\sum_{n=0}^\infty \frc{(it)^n}{n!}
	(\rep H)^n(A).
\eeq
It is easy to show that this is an absolutely converging power series with respect to the $C^\star$ algebra norm $||\cdot||$ with a nonzero radius of convergence. Let $\omega$ be a sizably clustering state. Then the series is also absolutely converging with respect to the norm on $\obsdh_\omega$, with, in general, a smaller nonzero radius of convergence, see Appendix \ref{appnonzero}.

In order to access large values of $t$, stronger results are needed. It is known that the continuation to $t\in\R$ makes $\tau_t:t\in\R$ into a strongly continuous one-parameter unitary group, and by Stone's theorem it is differentiable for all $t\in\R$ (see for instance \cite{BR1}). In particular, the following ``equations of motion'' hold
\beq\label{eom}
	\frc{\dd}{\dd t}\omega_t(A) = \ii\, \omega_t(\rep H(A)),\quad A\in\obs
\eeq
where the time-evolved state is $\omega_t:=\omega\circ\tau_t$; and $\tau_t$ preserves the algebra, $\tau_t(AB)=\tau_t(A)\tau_t(B)$ for all $A,B\in\obs$ (in both statements it is sufficient for us to restrict to local observables). We will also use the fact that $\tau_t$ preserves the trace state, $\Tr_\obsc(\tau_t(A)) = \Tr_\obsc(A)$ for all $A\in\obsc$.

Further, a result of Bravyi, Hastings and Verstraete \cite{BHV}, based on the Lieb-Robinson bound \cite{lr}, shows that $\tau_t(A)$ is in fact supported on a finite interval of length growing linearly with $|t|$, up to exponentially small corrections. The result uses natural projections $(\cdot)_X:\obs\to\obs_X$, $X\subset \Z^\di$ (whose precise form is not important here), and may be expressed as follows. For every local sequence $H$, there exist $\phi>0$, $\mu>0$ and $v>0$ such that for every $t\in\R$, every observable $A\in\obs$ and every subset $X\subset \Z^\di$,
\beq\label{timeexp}
	||\tau_t(A)-(\tau_t(A))_X||\leq \phi \,|A|\,||A||\,
	\exp\Big[-\mu\big(\dist(\supp(A),\Z^\di\setminus X)-v|t|\big)\Big].
\eeq
Let $X^A_n:=\cup_{x\in\ball(n)} \supp(\iso_x(A))$; this is a neighbourhood of $\supp(A)$ extending a distance $n$ from it. The bound \eqref{timeexp} shows that the projected time-evolved observables $(\tau_t(A))_{X_n^A}$ form for $n\in\N$ a Cauchy sequence with respect to $||\cdot||$, and
\beq\label{time2}
	\tau_t(A) = \lim_{n\to\infty} (\tau_t(A))_{X_n^A}.
\eeq

It turns out that, if $\omega$ is sizably clustering, then using \eqref{time2} we may also define $\tau_t(A)$ as an element of the Hilbert space $\obsdh_\omega$ (which of course agrees with \eqref{time} for $|t|$ small enough), and we may show that $\omega\circ\tau_t$ is sizably clustering.
\begin{theorem}\label{propotime}
Let $\omega$ be a sizably $p$-clustering state, and $\tau_t$ a time evolution as above.
\bi
\item[(i)] The state $\omega_t:=\omega\circ\tau_t$ is sizably $q$-clustering for every  $q<p$ and $t\in\R$, and it is so uniformly over any family $\{(\omega,t)\}$ where $t$ is in any compact subset of $\R$ and the states $\omega$ are uniformly sizably $p$-clustering.
\item[(ii)]  Let $A,B\in\obs$ and $t\in\R$. The limit \eqref{time2} exists in $\obsdh_\omega$, that is, $(\tau_t(A))_{X_n^A}$ form a Cauchy sequence with respect to $||\cdot||_{\obsdh_\omega}$. The resulting form is $\dbra\tau_t(A),\tau_t(B)\dket_\omega=\dbra A,B\dket_{\omega\circ \tau_t}$. The norm $||\tau_t(A)||_{\obsdh_\omega}$ is uniformly bounded, and the limit $\lim_{n\to\infty}||(\tau_t(A))_{X_n^A}||_{\obsdh_\omega}$ exists uniformly, over any family $\{(\omega,t)\}$ as above.
\ei
\end{theorem}
\proof
Assume that $\omega$ is sizably $p$-clustering. Consider the symmetric bilinear form
\beq
	(A,B)_\omega:=\omega(AB)-\omega(A)\omega(B).
\eeq
We first show that for all $q<p$ and all $t\in\R$ there exists $\nu_1,a_1>0$ (where $a_1$ is independent of $q$) such that
\beq\label{prtoshow1}
	|((\tau_t(A))_{X_n^A},B)_\omega|\leq \nu_1\ell^{a_1}\,
	||A||\,||B||\,\dist(A,B)^{-q}
\eeq
for all $n\in\N$, $\ell>0$ and $\siz{A},\siz{B}<\ell$. It is sufficient to show this for $\dist(A,B)$ large enough, and we will assume $\dist(A,B)>m_0$ where $m_0$ is the minimal integer such that $m_0 > (p/\mu)\log(m_0)+1$ (recall that $p$ is a property of the state $\omega$, and that $\mu$ is a property of the Hamiltonian $H$).

We will use the facts that $\dist(\supp(A),\Z^\di\setminus X_n^A) = n+1$ and $|X_n^A| \leq |\ball(n)|\,\siz{A}$. Let us set
\beq\label{prtm}
	m =(p/\mu)\log(\dist(A,B)).
\eeq
Clearly, $\dist(A,B)>m$. First assume that $n\leq m$. By sizable $p$-clustering of $\omega$, there exists $\nu_\omega,a_\omega\in\R^+$ such that
\beq\label{prtexp3}
	|((\tau_t(A))_{X_n^A},B)_\omega|\leq
	\nu_\omega (|\ball(n)|\ell)^{a_\omega}\,||(\tau_t(A))_{X_n^A}||\,||B||\,
	\dist(X_n^A,\supp(B))^{-p}
\eeq
for all $n\in\N$, $t\in\R$, $\ell>0$ and $\siz{A},\siz{B}<\ell$. Since $\tau_t$ is an isometry, it preserves the norm: $||\tau_t(A)||= ||A|| \;\forall\;A\in\obsc,\,t\in\R$. Using also the weaker form of \eqref{timeexp} given by $||\tau_t(A)-(\tau_t(A))_X||\leq \phi \,|A|\,||A||\,e^{\mu v|t|}$,
the following bound is obtained:
\beq\label{prtbounded}
	||(\tau_t(A))_{X}|| \leq (1+ \phi\siz{A}e^{\mu v |t|})
	||A||
\eeq
for every subset $X$. Further, since $\dist(A,B)>n$, we may write
\beq\label{prtdist}
	\dist(X_n^A,\supp(B))^{-p} \leq (\dist(A,B)-n)^{-p}
	\leq \dist(A,B)^{-p} (n+1)^p
	\leq \dist(A,B)^{-p} (2m)^p.
\eeq
We note that since $|\ball(n)|$ is a monotonically increasing polynomial in $n$ of degree $\di$, there exists $b>0$ such that $|\ball(n)|\leq (bn)^\di$ for all $n\geq 1$. Therefore, \eqref{prtexp3} gives
\beq\label{prtexp32}
	|((\tau_t(A))_{X_n^A},B)_\omega|\leq
	\nu_\omega\,\ell^{a_\omega} (1 + \phi\siz{A}e^{\mu v |t|})\, (bm)^{\di a_\omega} (2m)^p \, ||A||\,||B||\,
	\dist(A,B)^{-p}
\eeq
and thus for every $t\in\R$ there exists $\nu',a'\in\R^+$ such that
\beq\label{prqorf}
	|((\tau_t(A))_{X_n^A},B)_\omega|\leq
	\nu' \ell^{a'}\,m^{p+\di a_\omega} \,||A||\,||B||\,
	\dist(A,B)^{-p}
\eeq
for all $n\leq m$, $\ell>0$ and $\siz{A},\siz{B}<\ell$, with
\beq
	\nu' =\nu_\omega (1+\phi\,e^{\mu v|t|}) b^{\di a_\omega}2^p,\quad a'=a_\omega+1.
\eeq
Using the value \eqref{prtm} of $m$, the inequality \eqref{prqorf} can be recast into
\beq
	|((\tau_t(A))_{X_n^A},B)_\omega|\leq
	\nu'' \ell^{a'}\,||A||\,||B||\,
	\dist(A,B)^{-p}\,\,\big(\log(\dist(A,B))\big)^{p+\di a_\omega}
\eeq
with
\beq
	\nu'' = \lt(\frc{p}{\mu}\rt)^{p+\di a_\omega} \nu'.
\eeq
We write
\beq
	\dist(A,B)^{-p}\,\,\big(\log(\dist(A,B))\big)^{p+\di a_\omega}
	\leq \dist(A,B)^{-(p-\ep)}
	\sup_{d\geq 1} (d^{-\ep}(\log(d))^{p+\di a_\omega})
\eeq
where given any $\ep>0$ the supremum is finite. From this we conclude \eqref{prtoshow1} for all $n\leq m$, $\ell>0$, $q<p$ and $\siz{A},\siz{B}<\ell$, with $a_1=a_\omega+1$ and $\nu_1=\nu'''$ given by
\beq\label{prtnua}
	\nu''' = \nu_\omega 2^pb^{\di a_\omega}
	\lt(\frc{p (p+\di a_\omega)}{\mu e(p-q)}\rt)^{p+\di a_\omega} (1+\phi\,e^{\mu v|t|}).
\eeq

We now consider the case $n> m$. By \eqref{timeexp} and boundedness of $\omega$ we have
\beq\label{prtexp}
	|(\tau_t(A)-(\tau_t(A))_X,B)_\omega|\leq
	 2\phi \,|A|\,||A||\,||B||\,
	\exp\Big[-\mu\big(\dist(\supp(A),\Z^\di\setminus X)-v|t|\big)\Big]
\eeq
where the factor $2\phi$ is independent of $A$, $B$, $X$, $t$ and $\omega$.  Choosing $X=X_n^A$, this gives
\beq\label{prtexp2}
	|(\tau_t(A)-(\tau_t(A))_{X_n^A},B)_\omega|\leq
	 2\phi \,|A|\,||A||\,||B||\,
	e^{-\mu\big(n+1-v|t|\big)}.
\eeq
Therefore,
\beqa
	|((\tau_t(A))_{X_n^A},B)_\omega|
	&\leq&
	|((\tau_t(A))_{X_n^A}-(\tau_t(A))_{X_m^A},B)_\omega|
	+
	|((\tau_t(A))_{X_m^A},B)_\omega|\n
	&\leq&
	2\phi \,|A|\,||A||\,||B||\, 
	\lt(e^{-\mu\big(n+1-v|t|\big)} + e^{-\mu\big(m+1-v|t|\big)}\rt)
	+
	|((\tau_t(A))_{X_m^A},B)_\omega|\n
	&\leq&
	||A||\,||B||\,\lt( 4\phi \,|A|\,
	e^{-\mu\big(m+1-v|t|\big)}
	+
	\nu''' \ell^{a_1}\,\dist(A,B)^{-q}\rt)\label{prtexp5}
\eeqa
where in the last line we have used the result established for $n\leq m$ and $\nu'''$ is the constant \eqref{prtnua}. Using the value \eqref{prtm} of $m$, we have
\beqa
	|((\tau_t(A))_{X_n^A},B)_\omega|
	&\leq&
	||A||\,||B||\,\lt( 4\phi \,|A|\,e^{\mu v|t|}
	\dist(A,B)^{-p}
	+
	\nu''' \ell^{a_1}\,\dist(A,B)^{-q}\rt).\no
\eeqa
Since $p>q$ and $a_1>1$, this completes the proof of the full statement surrounding \eqref{prtoshow1}, with, explicitly, $a_1 = a_\omega+1$ and
\beq\label{prtmnu1}
	\nu_1 = \nu_1(q)= \nu'''+4\phi e^{\mu v|t|}
	= \nu_\omega 2^pb^{\di a_\omega}
	\lt(\frc{p (p+\di a_\omega)}{\mu e(p-q)}\rt)^{p+\di a_\omega} (1+\phi\,e^{\mu v|t|})
	+4\phi e^{\mu v|t|}.
\eeq

We now show that for all $q<p$ and all $t\in\R$ there exists $\nu_2,a_2>0$ (where $a_2$ is independent of $q$) such that
\beq\label{prtoshow2}
	|((\tau_t(A))_{X_n^A},(\tau_t(B))_{X_n^B})_\omega|\leq \nu_2 \ell^{a_2}\,
	||A||\,||B||\,\dist(A,B)^{-q}
\eeq
for all $n\in\N$, $\ell>0$ and $\siz{A},\siz{B}<\ell$. Again it is sufficient to show this for $\dist(A,B)$ large enough, and we will assume $\dist(A,B)>m_0$ as above. First in the case $n\leq m$ we may use \eqref{prtoshow1}, \eqref{prtbounded} (applied to $(\tau_t(B))_{X_n^B}$) and \eqref{prtdist} (applied to $\dist(X_n^B,\supp(A))$) to obtain, for all $q'<p$
\beq
	|((\tau_t(A))_{X_{n'}^A},(\tau_t(B))_{X_n^B})_\omega|\leq
	\nu_1(q') \ell^{a_1}\,(1 + \phi\siz{B}e^{\mu v |t|})\,
	(bm)^{\di a_1}(2m)^{q'}\,||A||\,||B||\,\dist(A,B)^{-q'}.
\eeq
Therefore, following the steps \eqref{prtexp32}-\eqref{prtnua}, we conclude  
\beq
	|((\tau_t(A))_{X_{n'}^A},(\tau_t(B))_{X_n^B})_\omega|\leq \nu'''' \ell^{a_2}\,
	||A||\,||B||\,\dist(A,B)^{-q}
\eeq
for all $n'\in\N$, $n\leq m$, $\ell>0$, $q<p$ and $\siz{A},\siz{B}<\ell$, with $a_2=a_1+1=a_\omega+2$ and
\beq\label{prtnua2}
	\nu'''' = \nu_1((p+q)/2) 2^{\frc{p+q}2}b^{\di a_1}
	\lt(\frc{(p+q) (p+q+2\di a_1)}{2\mu e(p-q)}\rt)^{\frc{p+q}2+\di a_1} (1+\phi\,e^{\mu v|t|})
\eeq
where we have chosen $q'=(p+q)/2$ (satisfying $q<q'<p$).
Taking $n'=n$ this leads to the statement \eqref{prtoshow2} in the case $n\leq m$, with $\nu_2=\nu''''$.

On the other hand, if $n>m$, we may reproduce the derivation \eqref{prtexp5} with $B$ replaced by $(\tau_t(B))_{X_n^B}$ by using the established case of \eqref{prtoshow2}, where $\nu''''$ appears in place of $\nu'''$ and $a_2$ in place of $a_1$. This give the full statement surrounding \eqref{prtoshow2}, with
\beq
	\nu_2 = \nu_2(q)=
	\nu_1((p+q)/2) 2^{\frc{p+q}2}b^{\di a_1}
	\lt(\frc{(p+q) (p+q+2\di a_1)}{2\mu e(p-q)}\rt)^{\frc{p+q}2+\di a_1} (1+\phi\,e^{\mu v|t|})
	+4\phi e^{\mu v|t|}.
\eeq

{\em Clustering of the state.} We use the above to conclude the first statement of the Theorem. Norm convergence of $(\tau_t(A))_{X_n^A}$ to $\tau_t(A)$ guarantees that $\lim_{n\to\infty}((\tau_t(A))_{X_n^A},(\tau_t(B))_{X_n^B})_\omega = (\tau_t(A),\tau_t(B))_\omega$. Therefore, taking the limit $n\to\infty$ in \eqref{prtoshow2}, the left-hand side tends to
\[
	|(\tau_t(A),\tau_t(B))_\omega|=|(A,B)_{\omega_t}|,
\]
and we have established that $\omega_t$ is sizably $q$-clustering for all $q<p$. Further, using the explicit forms of the bounding constants obtained from \eqref{prtnua} and \eqref{prtnua2}, we see that if a family of state $\{\omega\}$ is uniformly sizably $p$-clustering ($\nu_\omega\leq\nu,a_\omega\leq a\;\forall\;\omega$), then the family $\{\omega_t\}$ is uniformly sizably $q$-clustering for every $q<p$.

{\em Convergence in the Hilbert space.} We may use in addition $\tau_t(A)^\star = \tau_t(A^\star)$ to conclude the second statement of the Theorem as follows. By Theorem \ref{propoconv}, and since the clustering property \eqref{prtoshow1} is uniform over $n$, the summand in the series defining $\bra (\tau_t(A))_{X_n^A},B\ket_\omega$, which has a limit $n\to\infty$ pointwise, is uniformly bounded by a summable function (the right-hand side of \eqref{prtoshow1} with $A$ replaced by $\iso_x(A^\star)$). Therefore by Lebesgue's dominated convergence theorem the limit
\beq\label{prtlk}
	\lim_{n\to\infty} \bra (\tau_t(A))_{X_n^A},B\ket_\omega
\eeq
exists and is finite for every $A,B\in\obs$, and it does so uniformly over any family of uniformly bounded, uniformly sizably $p$-clustering states. Further, by \eqref{prtoshow2}, a similar argument shows that the limit
\beq\label{prtiapd}
	\lim_{n\to\infty} \bra (\tau_t(A))_{X_n^A},(\tau_t(B))_{X_n^B}\ket_\omega
\eeq
exists and is finite for every $A,B\in\obs$, and that it does so uniformly over any family of uniformly bounded, uniformly sizably $p$-clustering states. Taking $A=B$ in the latter, and since in \eqref{prtlk} we may take any $B\in\obs$ and $\obs$ generates a dense subset of $\obsh_\omega$, a standard application of the Reisz representation theorem implies that $n\mapsto (\tau_t(A))_{X_n^A}$ is a Cauchy sequence with respect to the norm in $\obsh_\omega$ for every $A\in\obs$. Similar arguments make $n\mapsto (\tau_t(A))_{X_n^A}$ into a Cauchy sequence with respect to the norm in $\obsh_\omega^\star$ for every $A\in\obs$, and thus also with respect to the norm in $\obsdh_\omega$. Further, by the above uniform boundedness, the norm $||(\tau_t(A))_{X_n^A}||_{\obsdh_\omega}$ is uniformly bounded, and $||(\tau_t(A))_{X_n^A}||_{\obsdh_\omega}$ converges uniformly to $||\tau_t(A)||_{\obsdh_\omega}$, on any family of uniformly bounded, uniformly sizably $p$-clustering states.

{\em Relation between Hilbert space and state.} Finally, observe that, by the above, the limit $n\to\infty$ exists pointwise on the summand in the infinite series definition of $\bra (\tau_t(A))_{X_n^A},(\tau_t(B))_{X_n^B}\ket_\omega$,
\[
	\lim_{n\to\infty}((\tau_t(\iso_x(A^\star))_{X_n^A},(\tau_t(B))_{X_n^B})_\omega = (\tau_t(\iso_x(A^\star)),\tau_t(B))_\omega = (\iso_x(A^\star),B)_{\omega_t}.
\]
By \eqref{prtoshow2}, with $\iso_x(A^\star)$ in place of $A$, the summand is uniformly bounded for $n\in\N$ and $x\in\Z^\di$ by a summable function of $x$. Hence the limit \eqref{prtiapd}, which is $\bra \tau_t(A),\tau_t(B)\ket_\omega$ by convergence in the Hilbert space, is given by that obtained by the pointwise limit, hence is equal to $\bra A,B\ket_{\omega\circ\tau_t}$. Similar arguments show $\dbra \tau_t(A),\tau_t(B)\dket_\omega = \dbra A,B\dket_{\omega\circ\tau_t}$.
\eproof

\begin{rema}
The existence of the continuous maps $\qmap$ and $\qmap^\star$ discussed in subsection \ref{rtquasi} (or the proof of Theorem \ref{propotime} itself) guarantees that the second statement in Theorem \ref{propotime} implies convergence of the limit with respect to the norm in $\obsh_\omega$ and $\obsh_\omega^\star$ as well.
\end{rema}

\begin{rema} Theorem \ref{propotime} shows that $\tau_t(A):=\lim_{n\to\infty}[(\tau_t(A))_{X_n^A}]^{\widehat{\;}}_\omega$ is an element of $\obsdh_\omega$ for every $A\in\obs$, where $[\cdot]_\omega^{\widehat{\;}}:\obs\to\obsdn_\omega$ is the coset map. It is therefore clear that $\tau_t(\cdot)$ is a linear map $\obs\to\obsdh_\omega$. However, it is not in general a linear map $\obsdn_\omega\to\obsdh_\omega$, as $\tau_t$ is not in general well defined on the coset space $\obs\setminus\obsdd_\omega$. Stronger conditions on $\tau_t$ are required for this, for instance the existence of $\gamma\in\R^+$ such that $|| \tau_t(A)||_{\obsdh_\omega} < \gamma || A||_{\obsdh_\omega}$ for all $A$.
\end{rema}

A simple corollary is the following:
\begin{corol} \label{corothermal} Let $(\beta,t_1,\ldots,t_k)\in R:=\{\beta\in\R:|\beta|<\beta_*\}\times \R\times\cdots\times \R$, and let $\omega_\beta^{\rm th}$ be a high-temperature Gibbs state associated to a local Hamiltonian and $\tau^{(1)}_{t_1},\ldots,\tau^{(k)}_{t_1}$ be time-evolution dynamics associated to possibly different local Hamiltonians. Consider the states $\omega_{\beta,t_1,\ldots,t_k}:=\omega_\beta^{\rm th}\circ \tau_{t_1}^{(1)}\circ\cdots\circ\tau_{t_k}^{(k)}$. Then the states $\omega_{\beta,t_1,\ldots,t_k}$ are sizably $p$-clustering for any $p>\di$, uniformly on any compact subsets of $R$.
\end{corol}
\proof This follows from Theorems \ref{theothermal} and \ref{propotime}.
\eproof

\begin{rema} \label{remaexpo}
By Theorem \ref{propotime}, a sizably exponentially clustering state $\omega$ with exponent $\alpha$ gives rise to states $\omega_t$ that are sizably $p$-clustering for all $p>\di$. In fact, it is a simple matter to adjust the proof to show that $\omega_t$ is also exponentially clustering, for all $t$ and for any exponent strictly smaller than $\alpha$. Corollary \ref{corothermal} can be likewise strengthened.
\end{rema}

We now show that pseudolocal states are well behaved under any finite time evolution.
\begin{theorem}\label{theowb} {\em (Stability of pseudolocal states)}
Let $\omega$ be a pseudolocal state, and $\tau_t$ a time evolution as above. Then $\omega\circ\tau_t$ is a pseudolocal state.
\end{theorem}
\proof
Let us denote by $q_s\in\obsdh_{\omega_s}$ the densities of the pseudolocal charges associated to the uniformly sizably $p$-clustering pseudolocal flow $\{\omega_s:s\in [0,1]\}$. We have $\omega_1=\omega$, and $\omega_0 = \Tr_{\obsc}$. Let $t>0$. First, by the fact that $\tau_t$ is an isometry, the family $\{\omega_s\circ\tau_t:s\in [0,1]\}$ is uniformly bounded, and by Theorem \ref{propotime}, it is uniformly sizably $q$-clustering for every $q<p$.

Let $A\in\obs$, and consider
\beq
	\omega_u(A) = \omega_0(A) + \int_0^u \dd s\,
	\dbra q_s,A\dket_{\omega_s}.
\eeq
Consider $\omega_u(\tau_t(A))$, obtained by the limit definition of $\tau_t(A)$ as in \eqref{time2}. Due to Theorem \ref{propotime}, the limit $\lim_{n\to\infty} \dbra q_s,(\tau_t(A))_{X_n^A}\dket_{\omega_s}$ exists pointwise in $s$. Observe that $||q_s||_{\obsdh_{\omega_s}}$ is uniformly bounded over $s\in[0,1]$. Observe also that, since the flow is uniform sizably $p$-clustering, then, by Theorem \ref{propotime}, the convergence of $||(\tau_t(A))_{X_n^A}||_{\obsdh_{\omega_s}}$ as $n\to\infty$ and the resulting limit is uniform on $s\in[0,1]$. Therefore, the integrand is uniformly bounded by an integrable function. Hence by Lebesgue's dominated convergence theorem, $\dbra q_s,\tau_t(A)\dket_{\omega_s}$ is bounded and Lebesgue integrable, and
\beq\label{prgge1}
	\omega_u(\tau_t(A)) = \omega_0(A) + \int_0^u \dd s\,
	\dbra q_s,\tau_t(A)\dket_{\omega_s}
\eeq
for all $u\in[0,1]$. In the first term on the right-hand side, we used the fact that the trace state is invariant under $\tau_t$.

Observe that $|\dbra q_s,\tau_t(A)\dket_{\omega_s}| \leq ||q_s||_{\obsdh_{\omega_s}}\,||\tau_t(A)||_{\obsdh_{\omega_{s}}} = ||q_s||_{\obsdh_{\omega_s}}\,||A||_{\obsdh_{\omega_{s}\circ\tau_t}}$, where the last equality is by Theorem \ref{propotime}. Thus, uniform boundedness of $||q_s||_{\obsdh_{\omega_s}}$ shows that the integrand in \eqref{prgge2} is a uniformly bounded continuous linear functional on $\obsdh_{\omega_{s}\circ\tau_t}$. Therefore by Theorem \ref{propotwosided} and the Reisz representation theorem, there exist uniformly bounded pseudolocal charges $\widehat Q_{\omega_{s}\circ\tau_t}$ such that, for every $A\in\obs$,
\beq\label{prgge3}
	(\omega_{u}\circ\tau_t)(A) = \omega_{0}(A) + \int_0^u \dd s\,
	\widehat Q_{\omega_{s}\circ\tau_t}(A).
\eeq
Hence, $\{\omega_{s}\circ\tau_t:u\in[0,1]\}$ is a uniformly sizably $q$-clustering pseudolocal flow from $\omega_0$ to $\omega\circ\tau_t$.
\eproof

Note that the time evolution of a weak pseudolocal state does not necessarily give rise to a weak pseudolocal state; uniform sizable $p$-clustering is required in the above proof.

Of course, infinite-time evolutions do not necessarily preserve the family of pseudolocal states. However, we show that under certain conditions they do give rise to a (weak) pseudolocal state, in which case the result is a (weak) GGE. This is expressed in the following theorem.
\begin{theorem} \label{theoGGE} {\em (Generalized thermalization)} Let $\tau_t$ be an evolution dynamics as above, and let $\omega$ be a pseudolocal state with flow $\{\omega_s:s\in[0,1]\}$. Suppose
\bi
\item[(a)] $\{\omega_s\circ \tau_t:(s,t)\in [0,1]\times [0,\infty)\}$ is uniformly clustering, and
\item[(b)] for every $A,B\in\obs$ and almost all $s\in[0,1]$, the limit $\lim_{t\to\infty} \dbra \tau_t(A),B\dket_{\omega_s}$ exists in $\C$.
\ei
Then the limit $\omega^{\sta}:=\lim_{t\to\infty}\omega\circ\tau_t$ exists ($\star$-weakly) and is a weak GGE with respect to the Hamiltonian $H$ of the dynamics. If there exists a $p>\di$ such that the clustering in (a) is uniformly sizably $p$-clustering, then $\omega^{\sta}$ is a GGE.
\end{theorem}
Note that by Theorem \ref{theowb}, the only nontrivial part of the requirement that  $\omega_s\circ \tau_t$ be uniformly clustering for $(s,t)\in [0,1]\times [0,\infty)$, is that uniformity hold over all large enough $t$.

\noindent\proof Let us denote by $q_s\in\obsdh_{\omega_s}$ the densities associated with the pseudolocal flow. Let $A\in\obs$. By the first part of the proof of Theorem \ref{theowb}, we have
\beq\label{prgge1}
	\omega_u(\tau_t(A)) = \omega_0(A) + \int_0^u \dd s\,
	\dbra q_s,\tau_t(A)\dket_{\omega_s}.
\eeq

Note that $|| \tau_t(A)||_{\obsdh_{\omega_s}} = ||A||_{\obsdh_{\omega_s\circ \tau_t}}$ by Theorem \ref{propotime}. By Theorem \ref{propoconv} and by the assumption of uniform  clustering of $\omega_s\circ\tau_t$, we find that $||\tau_t(A)||_{\obsdh_{\omega_s}}$ is uniformly bounded on $(s,t)\in [0,1]\times [T,\infty)$. Further, since the limit $\lim_{t\to\infty}\dbra \tau_t(A),B\dket_{\omega_s}$ exists for all $B\in\obs$ and almost all $s\in[0,1]$, an application of the Reisz representation theorem (using the fact that $\obsdn_{\omega_s}$ is dense in $\obsdh_{\omega_s}$) shows that, for almost all $s$, the limit $\tau_\infty(A)=\lim_{t\to\infty} \tau_t(A)$ exists in $\obsdh_{\omega_s}$. Therefore, the limit $t\to\infty$ exists pointwise in the integrand in \eqref{prgge1} except possibly on a set of measure zero. Since both $||q_s||_{\obsdh_{\omega_s}}$ and $||\tau_t(A)||_{\obsdh_{\omega_s}}$ are uniformly bounded on $s,t$, the integrand is bounded by an integrable function. Therefore, by Lebesgue's dominated convergence theorem, the limit $t\to\infty$ of the integral exists and is the integral of the pointwise limit; and the pointwise limit of the integrand, $\dbra q_s,\tau_\infty(A)\dket_{\omega_s}$, is bounded and Lebesgue integrable. Thus the limit $\omega_u^\sta(A):=\lim_{t\to\infty} \omega_u(\tau_t(A))$ exists for all $u\in[0,1]$ and $A\in\obs$, and gives
\beq\label{prgge2}
	\omega_{u}^\sta(A) = \omega_{0}(A) + \int_0^u \dd s\,
	\dbra q_s,\tau_\infty(A)\dket_{\omega_s}.
\eeq
Since, by definition of a pseudolocal flow and by the fact that the time evolution is an isometry (on $\obsc$), uniform boundedness holds on $(s,t)\in [0,1]\times [T,\infty)$, the limit $\omega_u^\sta$ is uniformly bounded on $u\in[0,1]$. Thus, $\omega_u^\sta$ can be extended to a uniformly bounded state on $\obsc$, and the $\lim_{t\to\infty}\omega_u\circ\tau_t$ exists weakly for every $u$ and gives $\omega_u^\sta$.

Note that uniform clustering of $\omega_s\circ\tau_t$ on $t$ implies that $\omega^\sta_s$ is clustering for all $s\in[0,1]$, and that, by Lebesgue's dominated convergence theorem, $\lim_{t\to\infty} ||A||_{\omega_s\circ\tau_t} = ||A||_{\omega_{s}^\sta}$ for all $s$. Since, then, $|\dbra q_s,\tau_\infty(A)\dket_{\omega_s}| \leq
||q_s||_{\obsdh_{\omega_s}}\,||\tau_\infty(A)||_{\obsdh_{\omega_{s}}} = ||q_s||_{\obsdh_{\omega_s}}\,||A||_{\obsdh_{\omega_{s}^\sta}}$, uniform boundedness of $||q_s||_{\obsdh_{\omega_s}}$ shows that the integrand in \eqref{prgge2} is a uniformly bounded continuous linear functional on $\obsdh_{\omega_{s}^\sta}$. Thus by Theorem \ref{propotwosided} and the Reisz representation theorem, there exist uniformly bounded pseudolocal charges $\widehat Q_{\omega_{s}^\sta}$ such that
\beq\label{prgge3}
	\omega_{u}^\sta(A) = \omega_{0}(A) + \int_0^u \dd s\,
	\widehat Q_{\omega_{s}^\sta}(A).
\eeq
Hence, $\{\omega_{u}^\sta:u\in[0,1]\}$ is a uniform clustering pseudolocal flow from $\omega_0$ to $\omega_{1}^\sta$.

Finally, by the equations of motion \eqref{eom}, we have $(\dd / \dd t) \, \omega_u(\tau_t(A)) = \ii \,\omega_u(\tau_t(\rep H(A)))$. Therefore both limits $\lim_{t\to\infty} \omega_u(\tau_t(A))$ and $\lim_{t\to\infty} (\dd/\dd t)\omega_u(\tau_t(A))$ exist in $\C$ for any $A\in\obs$. This implies that the derivative tends to 0, as we now show.

Let $f(t) = \omega_u(\tau_t(A))$, denote by $f'(t)$ its derivative and by $f_\infty\in\C$ and $f'_\infty\in\C$ the respective large-$t$ limits. Then the existence of the limits means that for all $\delta>0$ there exists a $T$ such that for all $u>t>T$, we have $|f(u)-f(t)|<\delta$ and $|f'(t)-f'_\infty|<\delta$. Therefore,
\[
	(u-t)|f'_\infty| \leq \lt|\int_t^u \dd s\,(f'(s)-f'_\infty)\rt|
	+\lt|\int_t^u \dd s\,f'(s)\rt|\leq (u-t)\delta + \delta
\]
and thus, taking $u$ arbitrarily large, $|f'_\infty|\leq \delta$. This holds for all $\delta>0$, which implies $f'_\infty=0$.

Therefore,
\beq
	\omega_{u}^\sta(\rep H(A))=0,
\eeq
which implies
\beq
	\int_I \dd s\,
	\widehat Q_{\omega_{s}^\sta}(\rep H(A)) = 0
\eeq
for every open interval $I\subset [0,1]$. Consider independently the real and imaginary parts of $\widehat Q_{\omega_{s}^\sta}(\rep H(A))$, and, for each, the set of values of $s$ on which it is positive and that on which it is negative. These are measurable sets because $\widehat Q_{\omega_{s}^\sta}(\rep H(A))$ is Lebesgue integrable (as a function of $s$). Since open intervals generate all Borel subsets, this implies that both the integrals of the positive parts and that of the negative parts are zero, and therefore $\widehat Q_{\omega_{s}^\sta}(\rep H(A))=0$ almost everywhere on $[0,1]$. For any $A\in\obs$, consider the zero-measure set $I_A$ on which $\widehat Q_{\omega_{s}^\sta}(\rep H(A))\neq 0$. Clearly $\obs$ has a countable basis, say $A_i\in\obs,\;i\in\N$. The countable union $\bigcup_{i\in\N}I_{A_i}$ is also of measure zero. Therefore, since $\widehat Q_{\omega_{s}^\sta}$ is linear, we find that for almost all $s\in[0,1]$, we have $\widehat Q_{\omega_{s}^\sta}(\rep H(A))=0$ for all $A\in\obs$.
\eproof

\begin{corol} {\em (Re-thermalization)} \label{theore}
Let $\beta\in \R$, and let $\{\omega_s=\omega_{s\beta}^{\rm th}:s\in[0,1]\}$ be a family of high-temperature Gibbs states with respect to some local Hamiltonian $G$ (in particular, $\beta$ must be small enough, a sufficient condition being $|\beta|<\beta_*$). Let $H$ be a (possibly different) completely mixing local Hamiltonian and $\tau_t$ the associated time evolution. Suppose
\bi
\item[(a)] $\{\omega_s\circ \tau_t:(s,t)\in [0,1]\times [0,\infty)\}$ is uniformly clustering, and
\item[(b)] for every $A,B\in\obs$ and almost all $s\in[0,1]$, the limit $\lim_{t\to\infty} \dbra \tau_t(A),B\dket_{\omega_s}$ exists in $\C$.
\ei
Then for every $s\in[0,1]$ the limit $\omega^{\sta}_{s\beta}:=\lim_{t\to\infty}\omega_{s\beta}^{\rm th}\circ\tau_t$ exists ($\star$-weakly), and is a weak GGE with respect to $H$. Further, suppose
\bi
\item[(c)] there is a neighborhood $K$ of $[0,1)$ such that for every $A\in\obs$, $\omega^{\sta}_{s\beta}(A)$ is analytic in $s\in K$ and uniformly clustering in every compact subset of $K$.
\ei
Then Statements (i) and (ii) of Theorem \ref{theoGGEgibbs} hold for the family $\{\omega_s=\omega_{s\beta}^{\sta}:s\in[0,1]\}$. In particular, under the conditions of the statements, $\omega^\sta_\beta$ is a high-temperature Gibbs state with respect to $H$, or an analytic extension thereof.
\end{corol}
\proof By Theorem \ref{theothermal}, $\omega_s$ is an analytic pseudolocal state for every $s\in[0,1]$. Conditions (a) and (b) of the theorem are directly conditions (a) and (b), respectively, of Theorem \ref{theoGGE}. Hence the limit $\omega_s^\infty:=\lim_{t\to\infty}\omega_s\circ\tau_t$ exists $\star$-weakly for every $s\in [0,1]$, and is a weak GGE with respect to $H$, whose flow (by the proof of Theorem \ref{theoGGE}) can be taken to be $\{\omega_{s'}^\infty:s'\in[0,s]\}$. Part (c) then guarantees that it is actually a weak analytic GGE, and thus Theorem \ref{theoGGEgibbs} can be used. \eproof

\begin{rema}
Theorem \ref{theoGGE} and Corollary \ref{theore} may be applied to ground states if boundedness and clustering holds uniformly all the way to zero temperature. This may hold, for instance, in gapped one-dimensional models, where it has been shown \cite{Hastings04,NS05} that the gap implies exponential clustering. However, one would need to check that clustering stays uniform in passing to the zero-temperature limit.
\end{rema}

\begin{rema}
In Theorems \ref{theoGGE} and Corollary \ref{theore}, one may replace $[0,\infty)$ by any unbounded sequence $t_n\in[0,\infty)$, and all limits on $t$ by limits on $n$, and the statements still hold.
\end{rema}

\section{Conclusion}\label{sectconclu}

In this paper we studied the concept of pseudolocality and its relation with thermalization. We first obtained various results concerning clustering states, forming the basis of the construction. Here clustering is to be strong enough, so that susceptibilities are finite (this includes exponential decay and strong enough algebraic decay of correlations). We then defined and studied pseudolocal sequences and the pseudolocal charges, which are linear functionals on observables, that these give rise to. We showed that pseudolocal charges are in bijection with the Hilbert space associated to susceptibilities, which can be interpreted as the space of their densities. We further defined a pseudolocal state as a state lying at the end-point of a flow whose tangents are determined by pseudolocal charges. These naturally generalize Gibbs states associated to local Hamiltonians. If the pseudolocal charges along the flow are all preserved by some dynamics, the pseudolocal state is defined to be a GGE. We showed that the family of pseudolocal states is preserved after finite time evolutions. Under certain conditions on the existence of the long-time limit, we showed that the pseudolocal state becomes a GGE in the limit of infinite time. The conditions involve a requirement of uniform clustering at long times, as well as the existence of the long-time limits of dynamical susceptibilities. Naturally, it would be useful to prove in explicit examples that the conditions of Theorem \ref{theoGGE} are fulfilled, but we believe the conditions are physically sound.

We note that clustering of the initial state was argued recently to play an important role in generalized thermalization under quadratic (free-particle) evolution Hamiltonians \cite{SC14}, and convergence to Gaussian states was recently shown rigorously \cite{GKFGE}. Also, it was found in \cite{So15} that the GGE description seems to generically fail for time evolutions controlled by Luttinger liquid Hamiltonians, as the ensuing ballistic transport precludes any thermalization. In this case, the condition of the existence of the large-time limit of dynamical susceptibilities is broken (thus this is in agreement with the present general theory).

The present results hold in any model, integrable or not. Integrability only plays a role in determining the space of conserved pseudolocal charges (thus the manifold of stationary states). We expect that in generic, non-integrable models, there be only a finite number of pseudolocal conserved charges, and in many cases the only one should be the Hamiltonian itself (that is, $H$ is completely mixing). In the latter case, as we showed, with some additional conditions, the associated GGE is a high-temperature Gibbs state, thus thermalization (in its usual sense) occurs.

It will be important to obtain a more in-depth understanding of the property of ``completely mixing", the space of conserved pseudolocal charges, and the relations with integrability. In particular, connecting the present general theory to the explicit conserved quasi-local charges constructed in the Heisenberg chain \cite{Pro14,PPSA14,IMP15,ZMP16} would be useful (for instance, do they form a basis for the associated Hilbert space?). Also, it would be interesting to connect the present re-thermalization result, Corollary \ref{theore}, to the thermalization theorem of \cite{MAMW15}; in the latter work, the condition of ``completely mixing'' is replaced by a condition of non-degeneracy on the spectra of finite-volume approximations of the evolution Hamiltonian.

As we have argued, the formal expression of GGEs ordinarily used in the literature agrees with the precise definition given here after certain formal manipulations (see Remark \ref{remaGGE}). In the usual treatment of GGEs, one assumes that all elements of the basis $Q_i$, used in the density matrix $e^{-\sum_i\beta_i Q_i}$, commute with each other.  However, there are situations where (pseudo)local conserved charges exist that do not commute with each other, and are involved in the GGE; this phenomenon was first pointed out in \cite{Fa14}, and recently it was discussed in the context of conformal field theory \cite{Ca15}. The definition in terms of flow still makes sense in such cases, indicating that, as we remarked, instead of exponentials in the formal expression one should use path-ordered exponentials.

We also note that an understanding of GGEs via equilibrium conditions could be useful: are GGEs, as here defined, KMS states associated to appropriate nonlocal interaction potentials? Are they maximal-entropy states? Do they satisfy a Gibbs condition?

As is clear in the analytic case, flows along pseudolocal charges suggest the notion an infinite-dimensional Riemannian manifold of clustering states, with metric tensor built out of the susceptibilities $\dbra\cdot,\cdot\dket_\omega$, and tangent space (a subspace of) that of pseudolocal charges. The submanifold generated by conserved pseudolocal charges may then be flat (if all pseudolocal charges generate commuting flows) or curved (otherwise), and a GGE would be the end-point of a flow lying on that manifold. The set of clustering states is not convex (contrary to the set of all states), and the infinite-dimensional manifold language seems to provide a potentially useful alternative framework.

A natural question is that of the approach to the GGE. It might be possible to develop and understanding of the emergence of hydrodynamics, and generalized versions of it associated to pseudolocal charges, in the present framework. A geometric interpretation via paths approaching of the conserved submanifold might also be fruitful. Other natural questions include that of constructing pseudolocal charges with respect to states with weaker clustering properties (such as critical ground states), convergence to GGEs from initial states with such weaker clustering, and / or weaker conditions on the large-time limit.

Pseudolocal states form a very wide family. In particular, we expect that states whose R\'enyi entropies scale like the volume of the system in a uniform enough fashion over the R\'enyi index (and satisfying other, related conditions) are pseudolocal states. It would be interesting to fully develop the relation between entanglement and pseudolocality. Combined with Theorem \ref{theoGGEgibbs}, this might then provide a proof of a version of the eigenstate thermalization hypothesis.

Finally, we hope that the framework provided in the present paper may be useful for other studies of non-equilibrium extended quantum systems.

{\bf Acknowledgements.}
I am grateful to Denis Bernard, Olalla A. Castro Alvaredo, Vincent Caudrelier, Jens Eisert, Toma$\check{\rm z}$ Prosen, Alexander Pushnitski and Spyros Sotiriadis for interest, discussions, and comments on the manuscript, as well as to the participants of the Isaac Newton Institute's programme ``Mathematical Aspects of Quantum Integrable Models in and out of Equilibrium" (11 January - 5 February 2016) for their comments and suggestions.

\appendix

\section{Proofs of Theorems \ref{propoclusform} and \ref{propocorr}}
\label{appform}

{\bf Proof of Theorem \ref{propoclusform}.} Let $A\in\obsh$. Since $\obsh$ is the completion of $\obs$, then $\obs$ is dense in $\obsh$, wherefore there exists a Cauchy sequence (with respect to the norm induced by $\bra\cdot,\cdot\ket$) $A_j\in\obs$ such that $\lim_{j\to\infty} A_j = A$. We first restrict to $B,C\in\obs$ such that $\omega(B)=\omega(C)=0$, and we fix some $\ell>0$. For every $j\in\N$, the form-clustering property \eqref{clusterform} implies that there exists $\nu_j>0$ such that
\beq
	|\bra A_j,BC\ket| \leq \nu_j\, ||B||\,||C||\,f(\dist(B,C)-\diam(A_j))
\eeq
whenever $\siz{B},\siz{C}<\ell$. By the Cauchy-Schwartz inequality, we also have, for any $D\in\obs$,
\beq\label{pr11}
	|\bra A_{j'},D\ket| \leq |\bra A_j,D\ket|
	+ ||A_{j'}-A_j||_\obsh \,||D||_\obsh.
\eeq
By the second part of Definition \ref{defiform}, there exists $\kappa>0$ such that $||BC||_\obsh\leq \sqrt{\kappa} \,||B||\,||C||$, and since $A_j$ is a Cauchy sequence, for every $\delta>0$ there exists a $j\in\N$ such that $||A_{j'}-A_j||_\obsh<\delta$ for all $j'\geq j$. Therefore, there exists $\kappa>0$ and for every $\delta>0$ there exists $j\in\N$, such that for every $\siz{B},\siz{C}<\ell$,
\beq
	|\bra A,BC\ket| \leq
	\lt(\nu_j\,f(\dist(B,C)-\diam(A_j))
	+ \delta \,\sqrt{\kappa}\rt)||B||\,||C||.
\eeq

Therefore for every $\ell,w>0$ we have $\lim_{\dist(B,C)\to\infty} |\bra A,BC\ket|\leq \delta\sqrt{\kappa} w^2$ uniformly on $B,C$ with $\siz{B},\siz{C}<\ell$ and $||B||,||C||<w$, and since this holds for every $\delta>0$, this shows that the limit is zero uniformly. In general, we may apply the result to $B-\omega(B)\1$ and $C-\omega(C)\1$, where we may use $||B-\omega(B)\1|| \leq ||B|| + |\omega(B)| \leq 2||B||$ and $\siz{B-\omega(B)\1} = \siz{B}$, and similarly for $C$. Recalling the fact that $\bra A,\1\ket=0$, this shows that the linear functional $\bra A,\cdot\ket$ satisfies part I of Definition \ref{defifunct}.

Let us denote $B_x:=\iso_x(B)$ and $C_x:=\iso_x(C)$. We now show that for all $B,C\in\obs$ such that $\omega(B)=\omega(C)=0$, there exists $\gamma>0$ such that for all $L\in\N$,
\beq\label{pr10}
	\sum_{x,y\in\ball(L)} |\bra B_x C,B_yC\ket| \leq \gamma L^{2\di-1}.
\eeq
Assume without loss of generality that $0\in\supp(B)$ and $0\in\supp(C)$, and let $\ell = \max\{\siz{B},\siz{C}\}$. We note that there exists $a\in\N$ such that $\diam(B_xC)\leq \dist(x,0)+a$ and that $\dist(B_x,C)\geq \dist(x,0)-a$ for all $x\in\Z^\di$. Thus, using the symmetry under $x\leftrightarrow y$, we may write
\beqa
	\sum_{x,y\in\ball(L)} \frc{|\bra B_x C,B_yC\ket|}{
	||B||^2\,||C||^2} &\leq&
	2\sum_{d,d'=0\atop d\geq d'}^{L} \sum_{x:\dist(x,0)=d\atop
	y:\dist(y,0)=d'} \frc{|\bra B_x C,B_yC\ket|}{
	||B||^2\,||C||^2}  \n
	&\leq& 2\nu(\ell) \sum_{d,d'=0\atop d\geq d'}^{L}
	\are_\di(d)\are_\di(d') f(d-d'-2a) \n
	&\leq& 2b^2 \nu(\ell)L^{2\di-2}\sum_{d=0}^{L}(L+1-d) f(d-2a) \n
	&\leq& 2b^2 \nu(\ell)L^{2\di-2}(L+1)\label{pfogj}
\eeqa
where in the third step we used the fact that there exists $b>0$ such that $\are_\di(d)\leq bL^{\di-1}$ (for all $d\in[0,L]$); and in the last step, we used summability over $d$ of $f(d-2a)$ ({\em a fortiori} from the fact that $d^{\di-1} f(d-2a)$ is summable). This shows \eqref{pr10}.

Eq. \eqref{pr10} implies $||\sum_{x\in\ball(L)} B_x C||_\obsh\leq \sqrt{\gamma} L^{\di-\frc12}$. We now use \eqref{pr11} with $D$ replaced by $\sum_{x\in\ball(L)} B_x C$, and thus for every $B,C\in\obs$ with $\omega(B)=\omega(C)=0$, we have
\beq
	\Big|\sum_{x\in\ball(L)}
	\bra A_{j'}, B_xC\ket\Big| \leq \sum_{x\in\ball(L)}  |\bra A_j,B_x C\ket|
	+ ||A_{j'}-A_j||_\obsh \,\sqrt{\gamma} L^{\di-\frc12}.
\eeq
Therefore, given $B,C$ and $\gamma$ as above, for every $\delta>0$ there exists $j\in\N$ such that
\beq
	\Big|\sum_{x\in\ball(L)}\bra A, B_xC\ket\Big|\leq 
	\t\nu_j\,\sum_{x\in\ball(L)} f(\dist(B_x,C)-\diam(A_j))
	+\delta \sqrt{\gamma} L^{\di-\frc12}
	\leq 
	\t{\t\nu}_j\,F
	\,\siz{B}\,\siz{C}
	+\delta \sqrt{\gamma} L^{\di-\frc12}
\eeq
where we define $\t\nu_j := \nu_j||B||\, ||C||$, and where in the second inequality we use an argument like that used in the proof of Theorem \ref{propoconv} (and the potentially different constant $\t{\t\nu}_j$ accounts for the part of the summation on negative arguments of $f$). The last inequality shows that the following limit exists and gives
\beq
	\lim_{L\to\infty} 
	L^{-\di+\frc12}\Big|\sum_{x\in\ball(L)}\bra A, B_xC\ket\Big| = 0
\eeq
for every $B,C$ as above. We obtain \eqref{clusterhalf} by applying the obtained result to $B-\omega(B)\1$ and $C-\omega(C)\1$ in place of $B$ and $C$.
\eproof

{\bf Proof of Theorem \ref{propocorr}.} Part II of the definition of the clustering form is immediate from Theorem \ref{propoconv}. Hence we concentrate on part I, and because of part II it is sufficient to show it for $\dist(B,C)>\diam(A)+1$ (so in particular $BC=CB$).

By Theorem \ref{propopos}, the sesquilinear form $\bra\cdot,\cdot\ket_\omega$ is positive semi-definite, and as noted it is translation invariant. We first show the statement of clustering of the form on observables with zero average: there exists $\t\nu(\ell)$ such that for every $\ell>0$ and every $A,B,C\in\obs$ with $\siz{A},\siz{B},\siz{C}<\ell$ and $\omega(A)=\omega(B)=\omega(C)=0$, we have \eqref{clusterform}, with $f(d)$ either algebraic or exponential.

Let
\beq\label{prk}
	v:=\diam(A)+1.
\eeq
We use the inequality (with the notation $A_x:=\iso_x(A)$)
\beq
|\bra A,BC\ket_\omega|
\leq  \sum_{x\in\Z^\di}| \omega(A_x^\star BC) |.
\eeq
Consider the following three subsets of $\Z$: $r_{AB,C}:=\{x\in\Z^\di:2\dist(A_x^\star B,C)\geq \dist(B,C)-v\}$, $r_{AC,B}:=\{x\in\Z^\di:2\dist(A_x^\star C,B)\geq \dist(B,C)-v\}$ and $r_{A,BC}:=\{x\in\Z^\di:2\dist(A_x, BC)\geq \dist(B,C)-v\}$. We are interested in the region formed by their union, $r = r_{AB,C}\cup r_{AC,B}\cup r_{A,BC}$. In this region, a bound can be established.

In the exponential case, we choose $0<\beta<\alpha/2$ and using state clustering we have
\beqa
	\lefteqn{
	\sum_{x\in r_{AB,C}\setminus r_{A,BC}}
	|\omega(A_x^\star B C)|} &&\n
	&\leq &
	e^{-\beta(\dist(B,C)- v)}
	\sum_{x\in r_{AB,C}\setminus r_{A,BC}}
	e^{2\beta\dist(A_x^\star B,C)}|\omega(A_x^\star B C)| \n
	&\leq& \nu(2\ell)\,||A||\,||B||\,||C||\,e^{-\beta(\dist(B,C)- v)}
	\sum_{x\in r_{AB,C}\setminus r_{A,BC}}
	e^{(2\beta-\alpha) \dist(A_x^\star B,C)}\n
	&\leq & \nu(2\ell)\,||A||\,||B||\,||C||\,e^{-\beta(\dist(B,C)- v)}
	\sum_{x\in r_{AB,C}\setminus r_{A,BC}}
	e^{(\beta-\alpha/2)( \dist(B,C)-v)}. \no
\eeqa
Likewise, using $BC=CB$,
\beqa
	\lefteqn{
	\sum_{x\in r_{AC,B}\setminus r_{A,BC}}
	|\omega(A_x^\star B C)|} &&\n
	&\leq &
	e^{-\beta(\dist(B,C)- v)}
	\sum_{x\in r_{AC,B}\setminus r_{A,BC}}
	e^{2\beta\dist(A_x^\star C,B)}|\omega(A_x^\star CB)|  \n
	&\leq & \nu(2\ell)\,||A||\,||B||\,||C||\,e^{-\beta(\dist(B,C)- v)}
	\sum_{x\in r_{AC,B}\setminus r_{A,BC}}
	e^{(\beta-\alpha/2)( \dist(B,C)-v)}. \no
\eeqa

In the algebraic case, we choose $\di<q<p-\di$ and using state clustering we have
\beqa
	\lefteqn{
	\sum_{x\in r_{AB,C}\setminus r_{A,BC}}
	|\omega(A_x^\star B C)|} &&\n
	&\leq &
	(\dist(B,C)- v)^{-q}
	\sum_{x\in r_{AB,C}\setminus r_{A,BC}}
	(2\dist(A_x^\star B,C))^q|\omega(A_x^\star B C)| \n
	&\leq& 2^p\nu(2\ell)\,||A||\,||B||\,||C||\,(\dist(B,C)- v)^{-q}
	\sum_{x\in r_{AB,C}\setminus r_{A,BC}}
	(2\dist(A_x^\star B,C))^{q-p}\n
	&\leq & 2^p\nu(2\ell)\,||A||\,||B||\,||C||\,(\dist(B,C)- v)^{-q}
	\sum_{x\in r_{AB,C}\setminus r_{A,BC}}
	(\dist(B,C)-v)^{q-p}\no
\eeqa
and
\beqa
	\lefteqn{\sum_{x\in r_{AC,B}\setminus r_{A,BC}}
	|\omega(A_x^\star B C)|} &&\n &\leq&
	2^p\nu(2\ell)\,||A||\,||B||\,||C||\,(\dist(B,C)- v)^{-q}
	\sum_{x\in r_{AC,B}\setminus r_{A,BC}}
	(\dist(B,C)-v)^{q-p}. \no
\eeqa

In both the exponential and algebraic cases, we have to evaluate the cardinals $|r_{AB,C}\setminus r_{A,BC}|$ and $|r_{AC,B}\setminus r_{A,BC}|$, times a function $g(\dist(B,C)-v)$ where $g(d)$ is finite and positive and decreases faster than $d^{-\di}$ as $d\to\infty$.

Let us analyze the cardinals of the summation sets $r_{AB,C}\setminus r_{A,BC}$ and $r_{AC,B}\setminus r_{A,BC}$. These are bounded by the cardinal of the set $\Z\setminus r_{A,BC}$. We are looking for all $x\in\Z^\di$ such that $2\dist(A_x,BC)<\dist(B,C)-v$. Since $\dist(B,C)>0$, we have that $\supp(B)\cap\supp(C)=\emptyset$, and so $\supp(BC) = \supp(B)\cup \supp(C)$ wherefore $\dist(A_x^\star,BC) = \min(\dist(A_x,B),\dist(A_x,C))$. Therefore we require either $2\dist(A_x,B)<\dist(B,C)-v$ or $2\dist(A_x,C)<\dist(B,C)-v$. There are at most $\ell^2 \are_\di(d)$ solutions $x$ to the condition $\dist(A_x,B)=d$, and also $\ell^2\are_\di(d)$ solutions to the condition $\dist(A_x,C)=d$, for any $d\in\N$. As a consequence, $|\Z\setminus r_{A,BC}|\leq \ell^2 \are_\di(\dist(B,C)-v) (\dist(B,C)-v)$, and we find
\beq
	|r_{AB,C}\setminus r_{A,BC}|\,g(\dist(B,C)-v),\ 
	|r_{AC,B}\setminus r_{A,BC}|\,g(\dist(B,C)-v)\quad
	\leq\quad \ell^2 \sup_{d>0} (d\are_\di(d)g(d)).
\eeq
Since $g(d)$ decays faster than $d^{-\di}$ and $\are_\di(d)$ is a polynomial in $d$ of degree $\di-1$, the supremum is (positive and) finite, $\sup_{d>0} (dg(d)) =:G<\infty$. Therefore,
\beq
	\sum_{x\in r_{AB,C}\cup r_{AC,B}\setminus r_{A,BC}
	}|\omega(A_x^\star B C)|\leq
	2^{p+1} \ell^2 G\,\nu(2\ell)\,||A||\,||B||\,||C||\,
	f(\dist(B,C)- v)
\eeq
where $f(d)=d^{-q}$ (algebraic case) of $f(d) = e^{-\beta d}$ (exponential case), where $G$ only depends on the powers or exponents of the clustering.

Moreover, we have
\beq
	\sum_{x\in r_{A,BC}}
	| \omega(A_x^\star BC) |
	\leq
	\nu(2\ell)\,||A||\,||B||\,||C||\,
	e^{-\beta (\dist(B,C)-v)}
	\sum_{x\in r_{A,BC}}
	e^{(2\beta-\alpha) \dist(A_x,BC)}
\eeq
(exponential case) and
\beq
	\sum_{x\in r_{A,BC}}
	| \omega(A_x^\star BC) |
	\leq
	\nu(2\ell)\,||A||\,||B||\,||C||\,
	(\dist(B,C)-v)^{-q}
	\sum_{x\in r_{A,BC}}
	\dist(A_x,BC)^{q-p}
\eeq
(algebraic case), and since $\siz{BC}\leq \siz{B}+\siz{C}<2\ell$, an argument like that used in Theorem \ref{propoconv} implies that we can bound the series by $2\ell^2$ times $\sum_{d=0}^\infty \are_\di(d) g(d)=:\t G<\infty$ for $g(d)$ as above, giving
\beq
	\sum_{x\in r_{A,BC}}
	| \omega(A_x^\star BC) |
	\;\leq\;2\ell^2\t G \,\nu(2\ell)\,||A||\,||B||\,||C||
	f(\dist(B,C)-v).
\eeq

Thus overall the summation is bounded in the region $r$ as per the statement of clustering of the form.

We now show that $r = \Z^\di$, and for this purpose we will show that $\Z^\di\setminus r$ is empty. In general, geometric considerations give
\beq\label{pr8}
	\dist(A_x,B)+\dist(A_x,C)+\diam(A)\geq \dist(B,C).
\eeq
Therefore if $\dist(A_x,B)=0$ then $\dist(A_x,C) \geq \dist(B,C)-\diam(A)$. From $\supp(A_x^\star B)\subset \supp(A_x)\cup\supp(B)$ the following inequality holds: $\dist(A_x^\star B,C)\geq\min(\dist(A_x,C),\dist(B,C))$. Therefore if $\dist(A_x,B)=0$ then $\dist(A_x^\star B,C)\geq \dist(B,C)-\diam(A)\geq (\dist(B,C)-v)/2$, using $v>\diam(A)$ from \eqref{prk}. Thus if $\dist(A_x,B)=0$ then $x\in r$. Similarly, if $\dist(A_x,C)=0$ then $x\in r$. Let us then assume that $\dist(A_x,B)\neq0$ and $\dist(A_x,C)\neq0$. In this case, since $\supp(A_x^\star B)= \supp(A_x)\cup\supp(B)$ we have $\dist(A_x^\star B,C)=\min(\dist(A_x,C),\dist(B,C))$, and similarly $\dist(A_x^\star C,B)=\min(\dist(A_x,B), \dist(B,C))$. Therefore, $\Z^\di\setminus r$ is contained in the set of solutions to the simultaneous conditions $2\dist(A_x,B)<\dist(B,C)-v$ and $2\dist(A_x,C)<\dist(B,C)-v$. Summing and using \eqref{pr8}, the set $\Z^\di\setminus r$ may be non-empty only if $\dist(B,C)-v>\dist(B,C)-\diam(A)$. With the choice \eqref{prk} for the value of $v$, the inequality is not satisfied, hence $\Z^\di\setminus r$ is empty.

The bound \eqref{clusterform} can then be established uniformly for all $A,B,C\in\obs$ with $\siz{A},\siz{B},\siz{C}<\ell$ by applying the established result to $A-\omega(A)\1$, $B-\omega(B)\1$ and $C-\omega(C)\1$; we use again $||A-\omega(A)\1|| \leq  2||A||$ and $\siz{A-\omega(A)\1} = \siz{A}$ (similarly for $B,C$), and $\bra \1,BC\ket_\omega=\bra A,\1\ket_\omega=0$.
\eproof

\section{Nonzero radius of convergence}\label{appnonzero}

We show that the series \eqref{time} has a nonzero radius of convergence in $\obsdh_\omega$.

Consider $\rep H^n(A)$. This is formed of terms of the type
\beq\label{prtype}
	h_{x_1}\cdots h_{x_j} A h_{x_{j+1}}\cdots h_{x_n}\quad\mbox{
	with coefficient 1 or $-1$}
\eeq
(where $h_x := \iso_x(h)$). Each such term has norm bounded by $||h||^n \,||A||$, and size bounded by $\siz{A}+n\siz{h}$. The number of such terms is constrained by the fact that in taking the commutator $\rep H(B)$, only the positions $x$ such that $\supp(h_x)\cap \supp(B)\neq \emptyset$ may occur, and each such position give rise to at most two terms of the type \eqref{prtype}. Thus the number of terms of the type \eqref{prtype} in $\rep H(B)$ is bounded by $2\,\siz{h} \,\siz{B}$. Let $K_n$ be an upper bound on the number of terms of type \eqref{prtype} in $\ad H^n(A)$. Then, by the above, we may set $K_{n+1} = 2\,\siz{h} \,(\siz{A}+n\siz{h})\, K_n $. Therefore $K_n = 2^n\siz{h}^n\,\prod_{m=0}^{n-1}(\siz{A}+m\siz{h})\leq 2^n\siz{h}^n  (\siz{A}+\siz{h})^n\,n!$. Assume that $\omega$ is sizably $p$-clustering. By Theorem \ref{propoconv}, using the triangular inequality, we have
\beq
	||\ad H^n(A)||_{\obsdh_\omega} \leq
	\sqrt{F \nu (\siz{A}+n\siz{h})^{a} }\,(\siz{A}+n\siz{h})\,||h||^n\,||A||\,
	2^n\siz{h}^n  (\siz{A}+\siz{h})^n\,n!.
\eeq
Therefore, the series \eqref{time} converges absolutely for all
\beq
	|t|< \lt(2 \siz{h}\,||h||\,(\siz{A}+\siz{h})\rt)^{-1}.
\eeq
\eproof

\end{document}